\documentclass[conference, letterpaper]{IEEEtran}

\IEEEoverridecommandlockouts
\ifCLASSINFOpdf
\else
\fi
  \usepackage{tipa}
  \usepackage{graphicx}
	\usepackage{epsfig}
	\usepackage{epstopdf}
	\usepackage{upgreek}
	\usepackage[nolist,nohyperlinks]{acronym}

\usepackage[cmex10]{amsmath}

\usepackage{color, soul}

\usepackage{subfigure}

\usepackage{cite}

\begin{acronym}
\acro{NoC}{Network-on-Chip}
\acro{WNoC}{Wireless Network-on-Chip}
\acro{EM}{Electromagnetic}
\acro{SiO$_2$}{Silicon Dioxide}
\acro{AIN}{Aluminum nitride}
\acro{SiP}{System-in-Package}
\acro{SDM}{Software-defined metamaterial}
\acro{MCM}{Multi-Chip Module}
\acro{TSV}{Through-Silicon Via}
\end{acronym}

\begin{document}



\title{Channel Characterization for Chip-scale Wireless Communications within Computing Packages
\thanks{The authors gratefully acknowledge support from the Spanish MINECO under grants PCIN-2015-012 and under contract TEC2017-90034-C2-1-R (ALLIANCE project) that receives funding from FEDER, from the EU's H2020 FET-OPEN program under grant No. 736876 (VISORSURF), and by the Catalan Institution for Research and Advanced Studies (ICREA).}
}


%
\author{
\IEEEauthorblockN{
Xavier Timoneda\IEEEauthorrefmark{1}, Albert Cabellos-Aparicio\IEEEauthorrefmark{1}, Dionysios Manessis\IEEEauthorrefmark{2}, Eduard Alarc\'{o}n\IEEEauthorrefmark{1}, Sergi Abadal\IEEEauthorrefmark{1}
}
\IEEEauthorblockA{
\IEEEauthorrefmark{1}NaNoNetworking Center in Catalunya (N3Cat), Universitat Polit\`{e}cnica de Catalunya (UPC), Barcelona, Spain\\%
\IEEEauthorrefmark{2}Fraunhofer Institute for Reliability and Microintegration (IZM), Berlin, Germany\\%
Email: xavier.timoneda@upc.edu, acabello@ac.upc.edu, dionysios.manessis@izm.fraunhofer.de,\\
eduard.alarcon@upc.edu, abadal@ac.upc.edu
}
}


\maketitle


\begin{abstract}
Wireless Network-on-Chip (WNoC) appears as a promising alternative to conventional interconnect fabrics for chip-scale communications. WNoC takes advantage of an overlaid network composed by a set of millimeter-wave antennas to reduce latency and increase throughput in the communication between cores. Similarly, wireless inter-chip communication has been also proposed to improve the information transfer between processors, memory, and accelerators in multi-chip settings. However, the wireless channel remains largely unknown in both scenarios, especially in the presence of realistic chip packages. This work addresses the issue by accurately modeling flip-chip packages and investigating the propagation both its interior and its surroundings. Through parametric studies, package configurations that minimize path loss are obtained and the trade-offs observed when applying such optimizations are discussed. Single-chip and multi-chip architectures are compared in terms of the path loss exponent, confirming that the amount of bulk silicon found in the pathway between transmitter and receiver is the main determinant of losses.
\end{abstract}



%
\IEEEpeerreviewmaketitle


\acresetall

\section{Introduction} \label{sec:introduction}
Recent years have witnessed a rising interest towards heterogeneous multi-chip architectures and the so-called 2.5D integration. The reasons are various, but mostly have their origin in the diminishing returns of transistor scaling and the cost of manufacturing large chips. In this context, heterogeneous multi-chip architectures allow to increase \emph{performance} of multicore processors beyond the limits of a large monolithic chip \cite{Arunkumar2017}; reduce their manufacturing \emph{cost} by disintegrating a large monolithic chip into a network of smaller ones, but with much better yield \cite{Kannan2016}; and provide \emph{versatility} or even \emph{modularity} in response to the increasing appeal of co-integrating diverse components such as CPUs, GPUs, memories, or accelerators within a single package \cite{Yin2018}. 
The new integration trends have strong consequences on the design of the communications backbone within the package. On the one hand, with the recent introduction of silicon interposers in 2.5D processes, there has been a reduction of the performance and cost difference between on-chip and off-chip communication \cite{Zhang2015d}. Also, interposers may be capable of hosting \emph{off-chip routers} in the future \cite{Kannan2016}. On the other hand, increased system heterogeneity implies higher versatility requirements as the actual communication needs will depend on the actual constituents of the architecture. 

It is expected that within-package networks will exploit the interposer advantages and rely on a tighter integration of the on-chip and off-chip sub-systems \cite{Kannan2016}. However, as off-chip transfers keep being expensive, it remains unknown whether such approach alone will suffice to meet the 
requirements of this scenario. As a result, emerging interconnect technologies are being explored as well \cite{Kim2012Survey, Thraskias2018}. Among them, wireless chip-scale communications 
show great promise due to its inherent lack of path infrastructure. This allows to overcome pin limitations and contribute to versatility by providing low-latency broadcast capabilities across the package \cite{Sujay2012}.

Research in highly-integrated wireless communications has exploded in the last decade \cite{Chen2009, Fettweis2013, Baniya2018a, Yu2014, Palesi2015, Abadal2018a, DiTomaso2015, Abadal2018, AbadalASPLOS, Choi2018}. The heterogeneous integration tendency has also impacted on this field, leading to several works of that explicitly consider wireless communications across chips for CPU-GPU coordination \cite{Gade2017a}, integrated memory access \cite{Sikder2016, Liu2016}, or in a more generic intra-/inter-chip framework \cite{Shamim2017}.

A missing piece in the wireless chip-scale puzzle is, however, the characterization of the wireless channel. The theory is well laid out \cite{Matolak2013CHANNEL} and a wide variety of works exist at on-chip \cite{Zhang2007, Rayess2017, Chen2018}, off-chip \cite{Chen2007prop, Narde2017, Baniya2018} and PCB board levels \cite{Chiang2010, Wu2013a, Kim2016mother}. However, as detailed in Section \ref{sec:related}, very few studies include the chip package in their simulations or measurements and, those that do it, are limited to low frequencies or lack proper justifications on the antenna type and placement \cite{Kim2001, Branch2005, Narde2018}. Note that, without proper understanding of the wireless channel within package, the path loss and dispersion assumptions may be overly optimistic. This affects the transceiver design and leads to inaccurate performance and efficiency reports. As most architectural studies rely in such figures, the impact of the wireless chip-scale paradigm cannot be really assessed.

\begin{figure*}[!t]
\centering
\subfigure[3D integration\label{fig:3D-int}]{\includegraphics[width=0.26\textwidth]{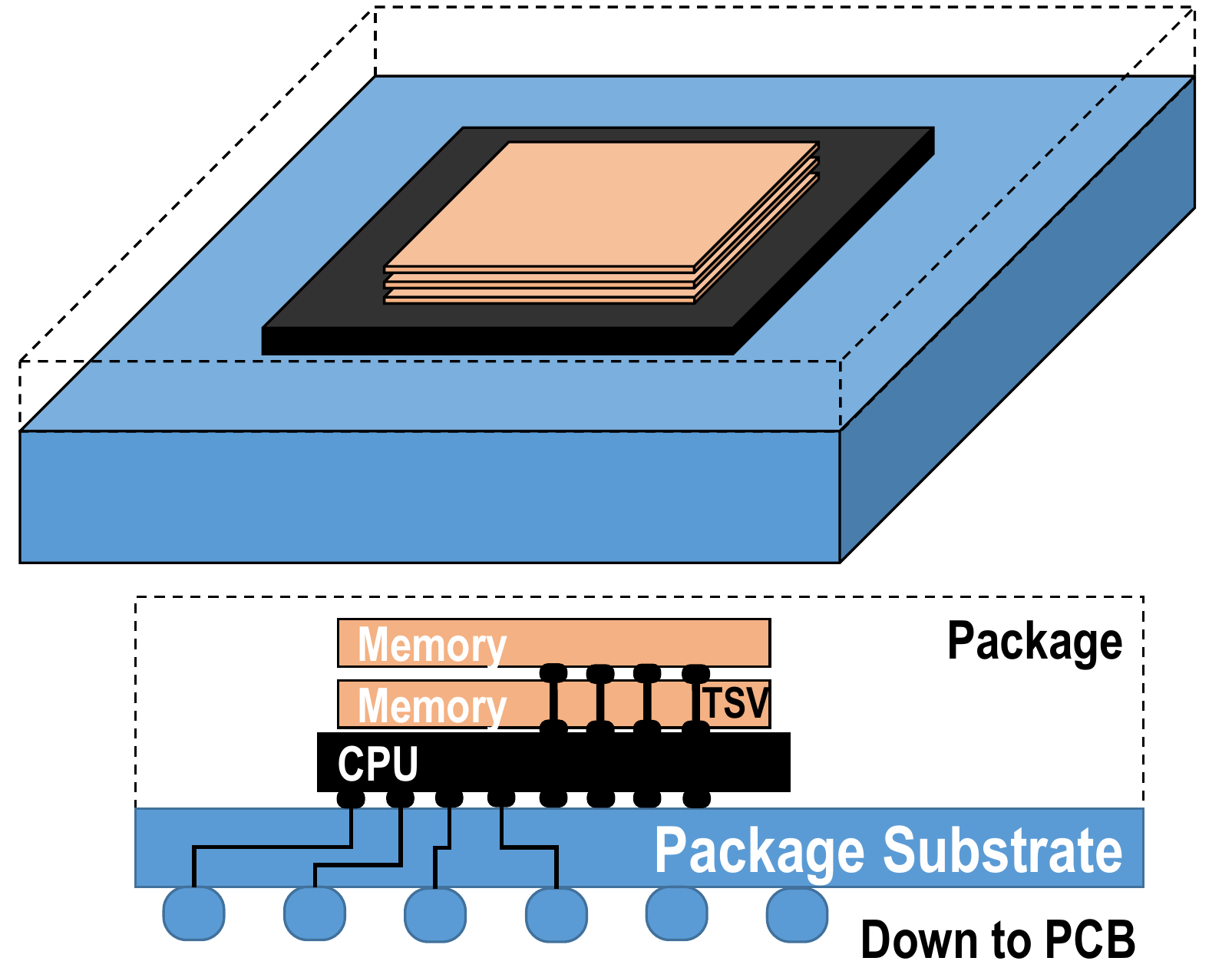}}
\subfigure[Silicon interposer\label{fig:interposer}]{\includegraphics[width=0.27\textwidth]{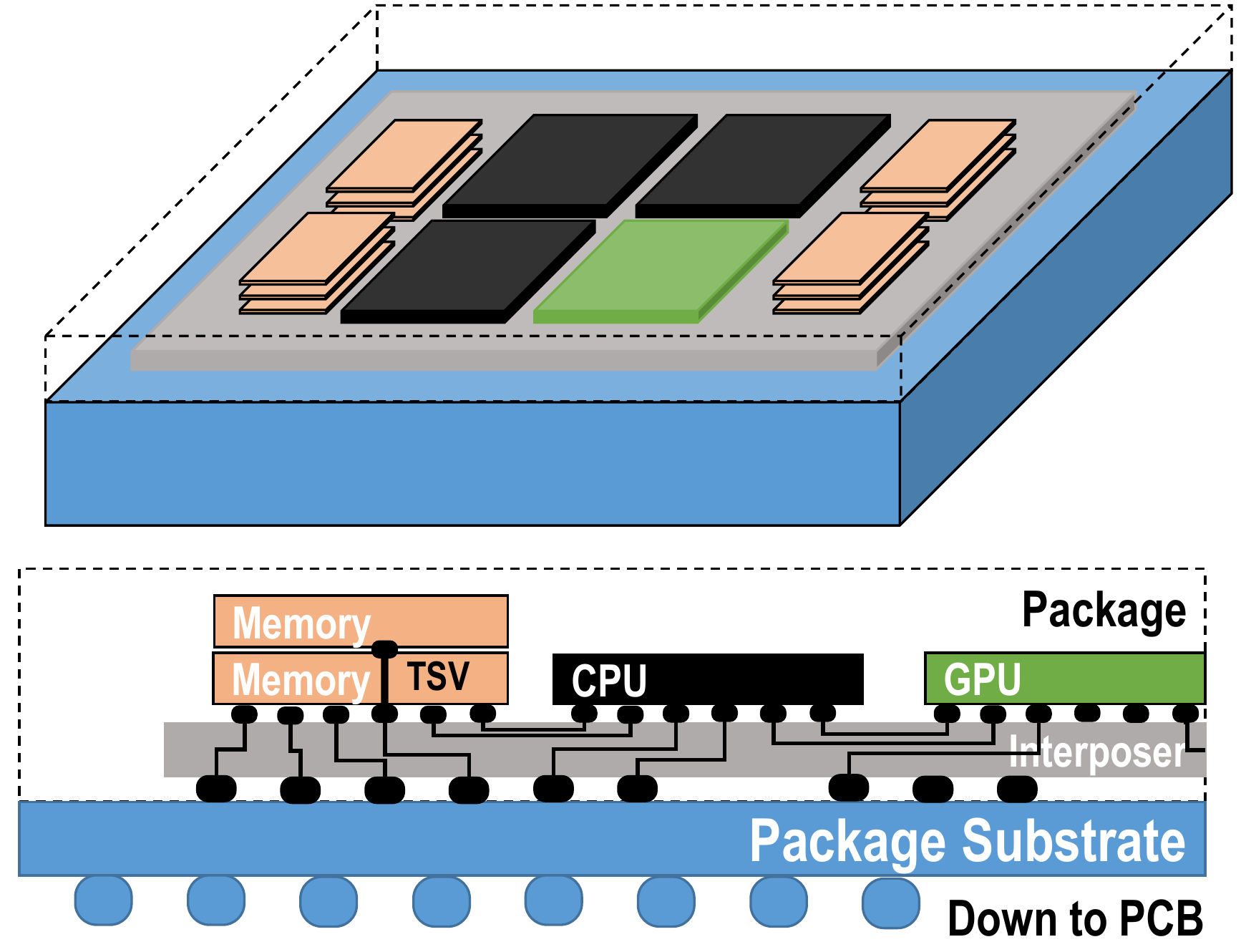}}
\subfigure[Multi-chip module\label{fig:MCM}]{\includegraphics[width=0.4\textwidth]{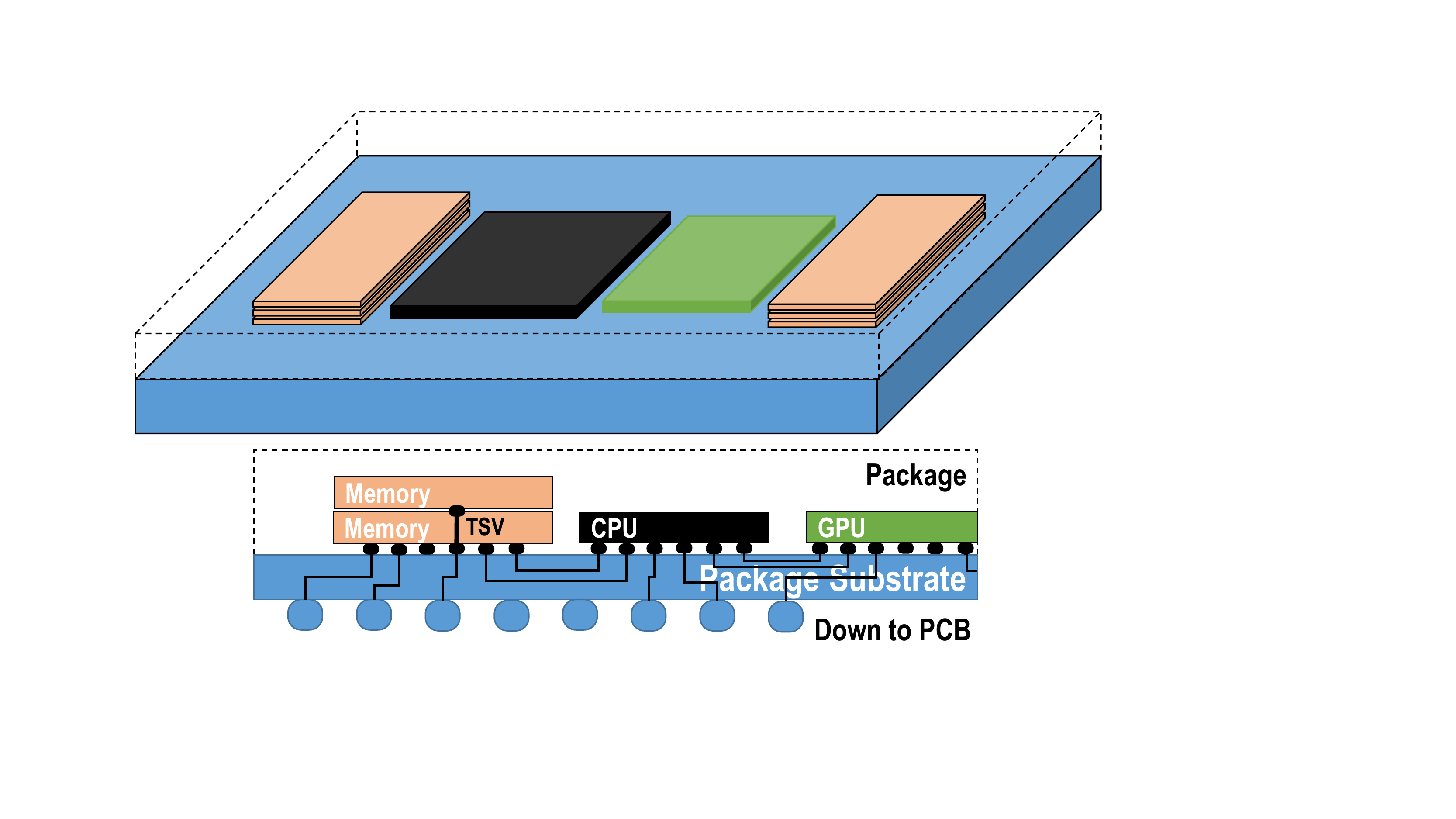}}
\vspace{-0.2cm}
\caption{Different heterogeneous integration techniques in examples with CPUs, GPUs, and memory. In a wireless approach, selected components would be equipped with one or several integrated antennas for wireless communication within the package.}
\label{fig:package}
\end{figure*} 

This paper aims to address this issue by providing a characterization of wireless channels compatible with heterogeneous 2.5D packaging. We rigorously model the package in a variety of single-chip and multi-chip configurations and discuss the antenna placement. By means of full-EM simulation, we extract the field distribution and coupling between antennas and then derive path loss models. Through parametric studies, we obtain optimal package dimensions for path loss minimization and also analyze the impact of the die-package transitions in multi-chip configurations. Although the methodology is applicable to any antenna type and frequency range, we particularize it for the promising case of \acp{TSV} used as monopoles in the range of 50--150 GHz. 

The remainder of the paper is as follows. Section \ref{sec:system-model} presents the system model, including details of the chip package and main assumptions. Section \ref{sec:methodology} describes the simulation methodology and subsequent channel modeling, whereas Section \ref{sec:simulationresults} presents the main results. Finally, Section \ref{sec:related} analyzes related works and Section \ref{sec:conclusion} concludes the paper.

\section{System Model}
\label{sec:system-model}
This work considers a variety of multi-chip configurations for the channel characterization, summarized in Figure \ref{fig:package}. All cases assume flip-chip integration. Although heat dissipation schemes are generally applied on a per-chip basis, here we propose the addition of a single heat spreader common to all chips and then the heat sink on top. Next, we provide more details on the structure of the package, dimensions, and materials. 

\subsection{Multi-chip Integration} 
\label{sec:multi-chip}
The (heterogeneous) integration of multiple chips currently takes place either vertically or horizontally. The former, represented in Figure \ref{fig:3D-int}, consists on the stacking of several chips that have been previously thinned down below 100 $\upmu$m \cite{Topol2006}. Once stacked, the chips are interconnected through a forest of vertical \acp{TSV} with very fine pitch. This provides a huge bandwidth density and
efficiency due to the very short link lengths. On the downside, 3D integration suffers from evident heat dissipation issues and the available area of integration basically depends on the dimensions of the chip at the base, i.e. around 20$\times$20 mm\textsuperscript{2}. 
 
Contrary to 3D stacking, heterogeneous 2.5D integration takes a co-planar approach and interconnects chips either through a common platform \cite{Zhang2015d}. Depending on the level of integration, this common platform may be silicon interposer (Fig. \ref{fig:interposer}) or the package substrate in a more classical \ac{MCM} approach (Fig. \ref{fig:MCM}). Such an arrangement alleviates the heat dissipation issue of 3D stacking and also increases the available area, as the limit is now set by the interposer (24$\times$36 mm\textsuperscript{2} in \cite{Kannan2016}, 40$\times$40 mm\textsuperscript{2} in \cite{Zhang2015d}) or the substrate (77$\times$77 mm\textsuperscript{2} in \cite{Arunkumar2017}). It also reduces the cost of the interconnects, as the pitch of \acp{TSV} is significantly coarsened. The main downturn of the approach is the reduction of bandwidth density and efficiency due to pin limitations and the need for longer links.

As for heat dissipation, it is worth noting that heat dissipation schemes are generally applied to each chip individually and then covered by a common lid. Instead, in his work we propose the addition of a single heat spreader common to all chips of a multi-chip configuration, and then a single heat sink on top. This would enhance heat dissipation further and favor inter-chip propagation through a common layer, reducing losses due to reflections at the chip-package interfaces. Molding compounds are sometimes used to fill the gaps between chips and below the heat spreader \cite{ardebili2009encapsulation}. However, due to its poor thermal behavior, we advocate to the direct interfacing of the chip with the heat spreader. The lateral space between chips is assumed to be filled with air or vacuum.
  
This paper explicitly considers wireless communication in 2.5D environments. To this end, we model the interposer and \ac{MCM} cases and comparing them with single-chip architectures. Therefore, the 3D stacking case is also indirectly represented: a single-chip architecture with thin silicon can be seen as a 3D stack as long as the antenna is placed on the top layer, just before the heat spreader.


\subsection{Flip-chip Package}
Although a recent work suggests a packageless architecture \cite{Pal2018}, dies have historically included a package to (i) act as a space transformer for I/O pins, provide mechanical support to the dies, and (iii) for ease of testability and repairability. Some packages include a molding compound around the chip to improves mechanical stability \cite{ardebili2009encapsulation}, but its typically poor thermal conductivity discourages its use in hot architectures. In most cases, even the packageless one \cite{Pal2018}, the die can be contacted directly by a Thermal Interface Material (TIM) with a metallic heat sink on top, avoiding the use of the molding compound. 

This work considers a flip-chip package with solder bumps. The packaging procedure is summarized here; we refer the reader to \cite{Timoneda2018} and references therein for more details. During the manufacturing process, the solder bumps are deposited on the chip pads, which already carry a valid under bump metallization (UBM) like nickel/gold (Ni/Au). Then, the chip is flipped over and its solder bumps are aligned precisely to the pads of the package carrier external circuit. This is in contrast to wire bonding of chips on the package substrate or the interposer, in which the chip is mounted upright and wires are used to interconnect the chip pads to external circuitry \cite{Branch2005}. Flip chip is generally preferred over wire bonding due to (i) its much lower inductance given by the shorter interconnect length \cite{Wright2006}, (ii) lower power--ground inductance due to direct routing of power, and (iii) higher power density given by the use of the whole chip surface. 

\begin{figure}[!t]
\centering
\includegraphics[width=0.9\columnwidth]{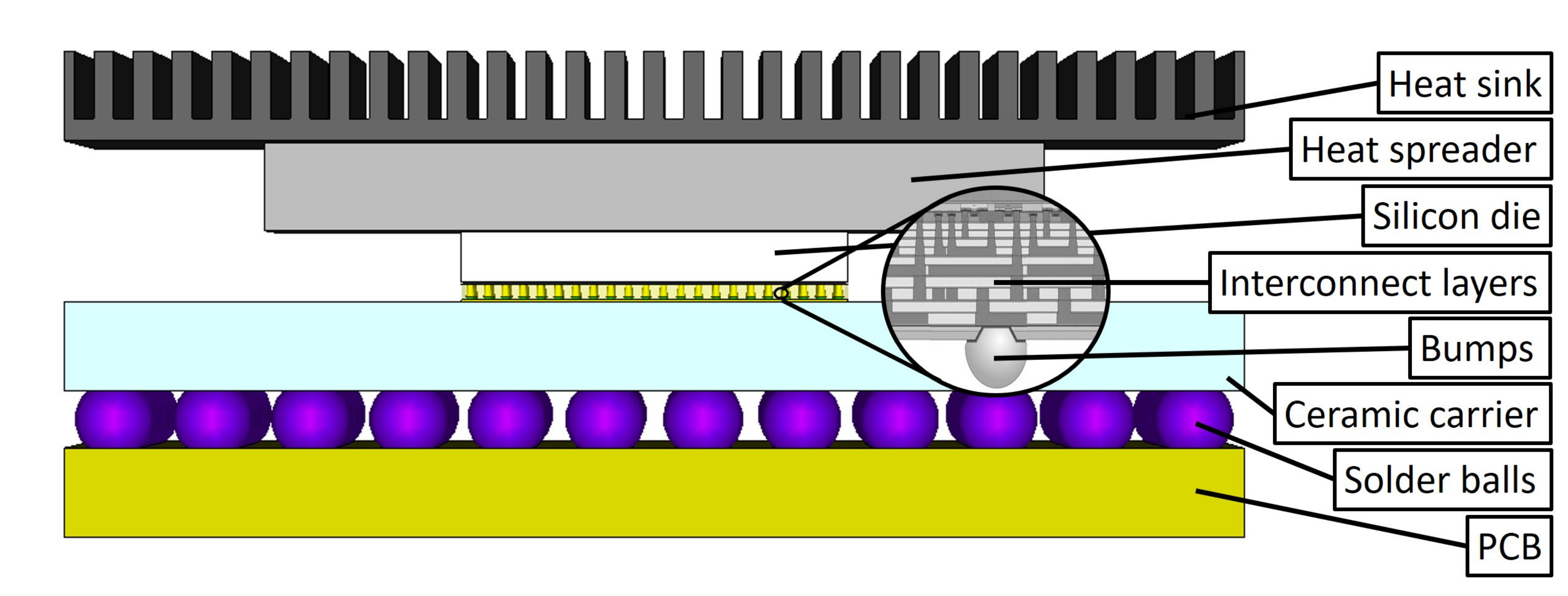}
\caption{Schematic of the layers of a flip-chip package}
\label{fig:flipchip}
\end{figure}

An instance of the resulting complete package is shown in Figure \ref{fig:flipchip}. The layers are described from top to bottom as summarized in Table~\ref{tab:flipchip}. On top, the heat sink and heat spreader dissipate the heat out of the silicon chip, as they both have good thermal conductivity. Bulk silicon serves as the foundation of the transistors. This layer has low resistivity (10 $\Upomega\cdot$cm), which is convenient for the operation of transistors, but not for electromagnetic propagation \cite{Kimoto2009}. The interconnect layers, which occupy the bottom of the silicon die as shown in the inset of Fig. \ref{fig:flipchip}, are generally made of copper and surrounded by an insulator such as silicon dioxide (SiO$_2$) \cite{Markish2015}. Depending on the case, we find a silicon interposer or a package substrate below the micro-bumps.

\begin{table}[!t] 
\caption{Characteristics of the layers in a computing package}
\vspace{-0.1cm}
\label{tab:flipchip}
\footnotesize
\centering
\begin{tabular}{lcccc} 
\hline
& {\bf Thickness} & {\bf Material} & {\bf $\varepsilon_{r}$} & tan($\delta$) \\
\hline
Heat sink & 0.5 mm & Aluminum & - & - \\
Heat spreader & 0.8 mm & Thermal cond. & 8.6 & 3$\cdot$10\textsuperscript{-4}\\
Silicon die & 0.2 mm & Bulk Silicon & 11.9 & 0.2517 \\
Interconnections & 13 $\upmu$m & Cu and \textbf{SiO$_2$} & 3.9 & 0.03\\
Bumps & 87.5 $\upmu$m & Cu and Sn & -  & - \\
Interposer & 0.1 mm & Bulk Silicon & 11.9 & 0.2517 \\
Ceramic carrier & 0.5 mm & Alumina & 9.4 & 4$\cdot$10\textsuperscript{-4}\\
\hline
\end{tabular}
\vspace{-0.5cm}
\end{table}

\subsection{Package Optimization for EM Propagation}
\label{sec:opti}
In our previous work, we discussed the impact of the different materials of a chip package on electromagnetic propagation. As pointed out above, the chip substrate and the heat spreader are the main determinants of the path loss and, by modifying their thicknesses, we can optimize propagation. 

\begin{figure}[!t]
\centering
\subfigure[Improvement over 0.7-mm Si\label{fig:gainvssi60}]{\includegraphics[width=0.48\columnwidth]{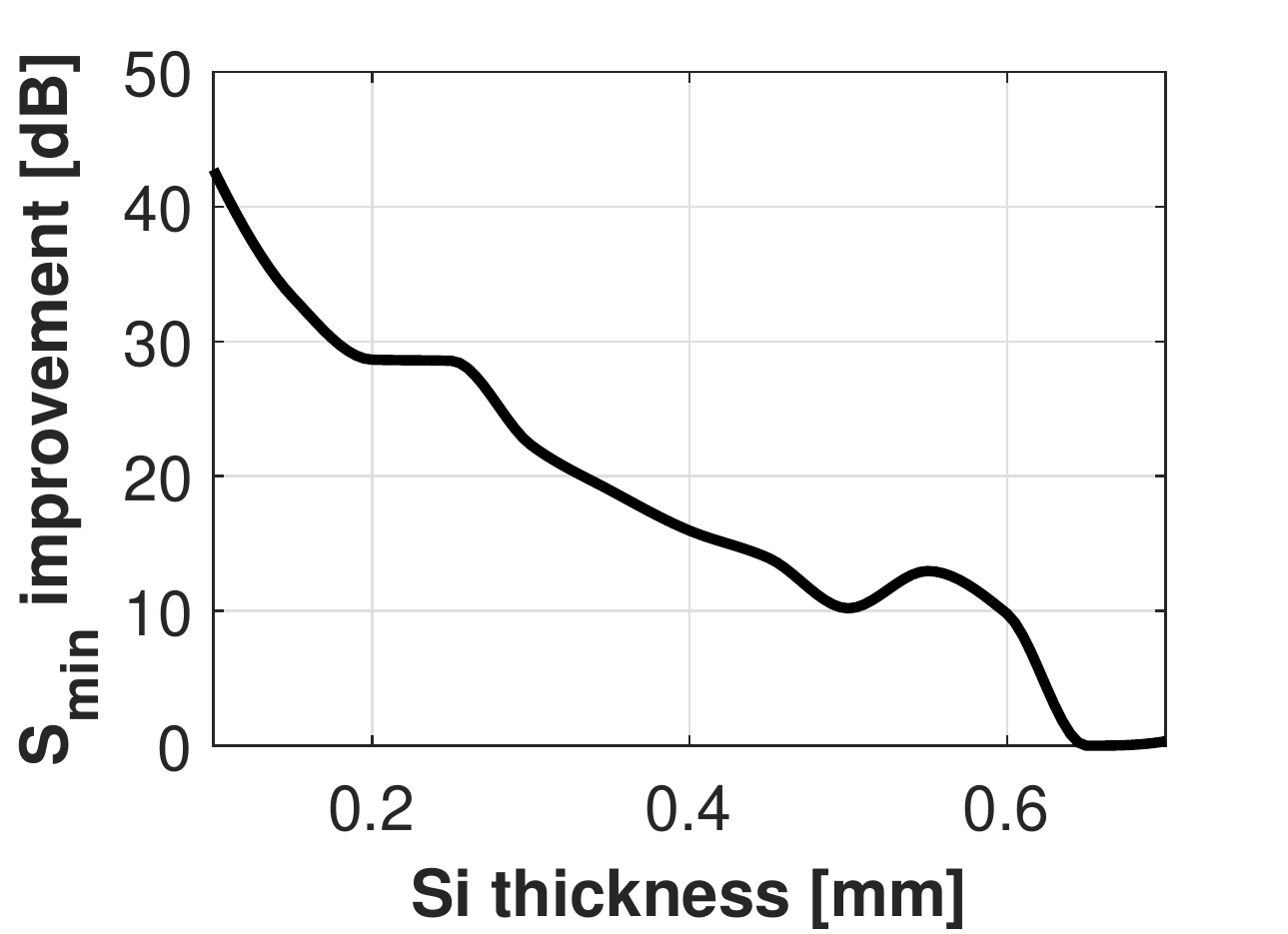}}
\subfigure[Improvement over no AIN\label{fig:gainvsain60}]{\includegraphics[width=0.48\columnwidth]{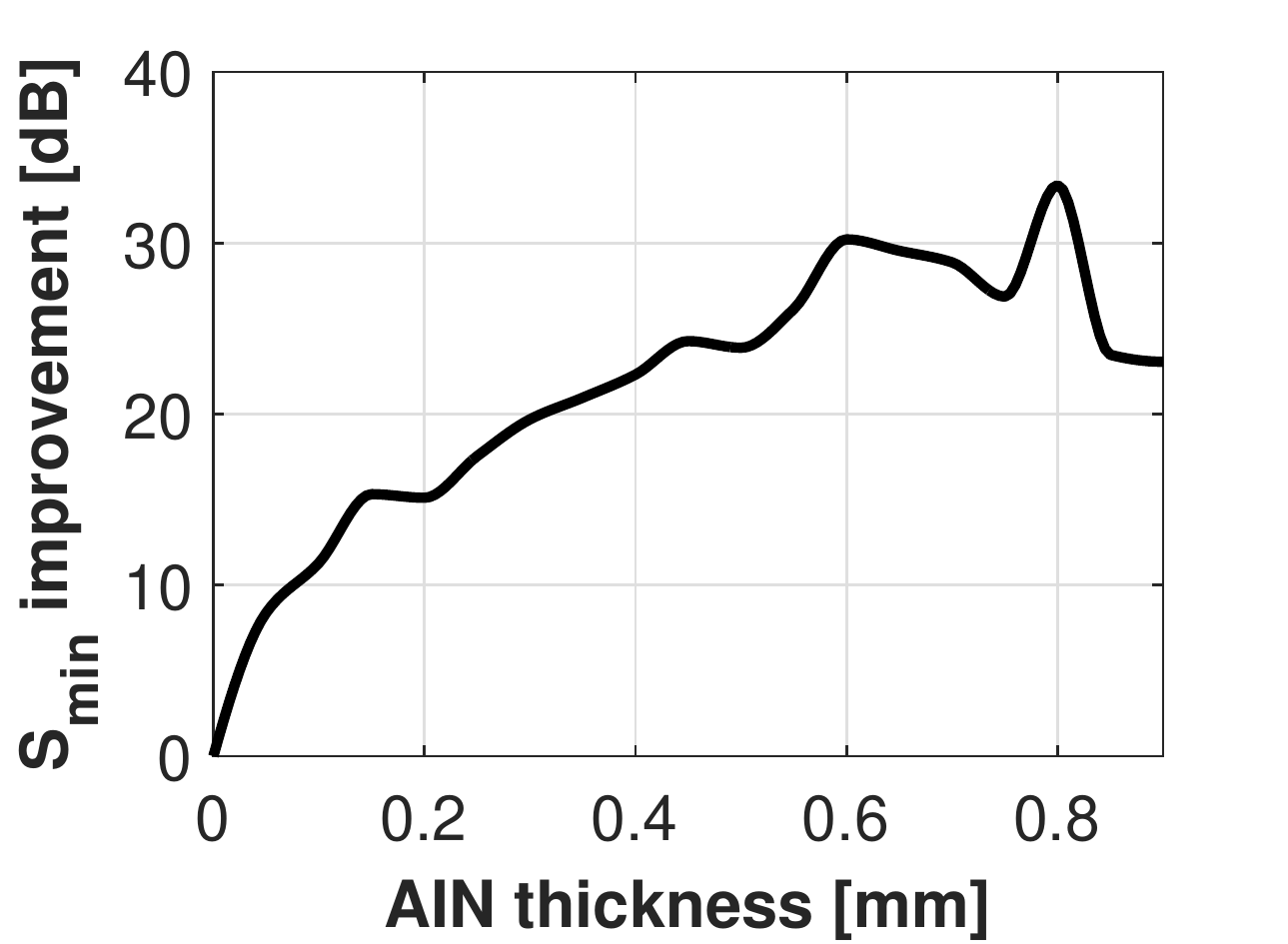}}
\caption{Adjusting the Silicon and heat spreader thickness result in huge improvements in the path loss.}
\label{fig:gainvsainsiold}
\end{figure}

The bulk silicon used in the chip substrate generally has low resistivity, which means a high loss tangent, and therefore we proposed to thin it to minimize propagation at this layer. To quantify the gains of this process, in \cite{Timoneda2018} we studied the path loss for different silicon thickness values in a single-chip package. We took 100 $\upmu$m as lower limit, frequently assumed in 3D stacking, although chip makers can reportedly reduce that further to tens of microns \cite{Bieck2010}. As we can see in Fig. \ref{fig:gainvssi60}, the path loss difference between the 0.1-mm and 0.7-mm cases is over 40 dB. Henceforth, we take 200 $\upmu$m as the value by default.

The materials used as heat spreaders have good thermal properties and, coincidentally, low electrical losses \cite{Kimoto2009}. To study their potential impact on electromagnetic propagation, in \cite{Timoneda2018} we simulated a chip package with different heat spreader thicknesses --our choice was Aluminum Nitride (AIN). As observed in Fig. \ref{fig:gainvsain60}, thickening the heat spreader reduces losses up to 33 dB with respect to not having any heat spreader. Therefore, it is a parameter to consider in package engineering efforts. Henceforth, we consider a 800 $\upmu$m AIN layer as the heat spreader by default.

\subsection{Antenna Integration}
The antenna placement within a flip-chip package is another important design consideration. Placing the radiating element as far from the lossy silicon as possible, as proposed in several works \cite{Zhang2005, Zhang2007, Gutierrez2009}, is not realistic because the antenna would be short-circuited by the array of micro-bumps. Instead, printed dipoles \cite{Branch2005} or patch antennas \cite{Yordanov2016} may be implemented in the metal layers closest to the silicon. However, the proximity of the antennas to the \emph{virtual ground plane} formed by the array of micro-bumps reduces their efficiency, whereas co-planarity between antennas further increases losses.

Finally, one could use \acp{TSV} as quarter-wave monopole antennas for several reasons: (i) the antenna would radiate laterally, directly towards the receiving antennas; (ii) advanced TSV and electroplating techniques \cite{FraunhoferTSV} would allow fine-tuning the antenna to the desired frequency; and (iii) the array of micro-bumps would naturally act as a ground plane, allowing to see the quarter-wave monopole effectively as a dipole. Note that vertical on-chip monopoles have been proposed recently \cite{Wu2017b}, but using non-standard fabrication and packaging.

Given the promising performance of monopoles in the chip-scale environment, we will consider them throughout this work. Figure \ref{fig:monopole} shows a sketch together with the expected radiation pattern within the package.

\section{Methodology}
\label{sec:methodology}
The canonical structure of Fig. \ref{fig:flipchip} is introduced in CST MWS \cite{CST} with the parameter values from Table \ref{tab:flipchip}. We then modify the structure to model the different scenarios depicted in Fig. \ref{fig:package} and to perform package optimizations. To reduce the computational burden, we perform several approximations that do not affect the accuracy of the results. For instance, given their fine-grained pitched at mm-Wave frequencies, the micro-bump array placed between the chip and the package substrate/interposer is modeled as solid metallic element \cite{Timoneda2018}.

\begin{figure}[!t]
\centering
\includegraphics[width=0.75\columnwidth]{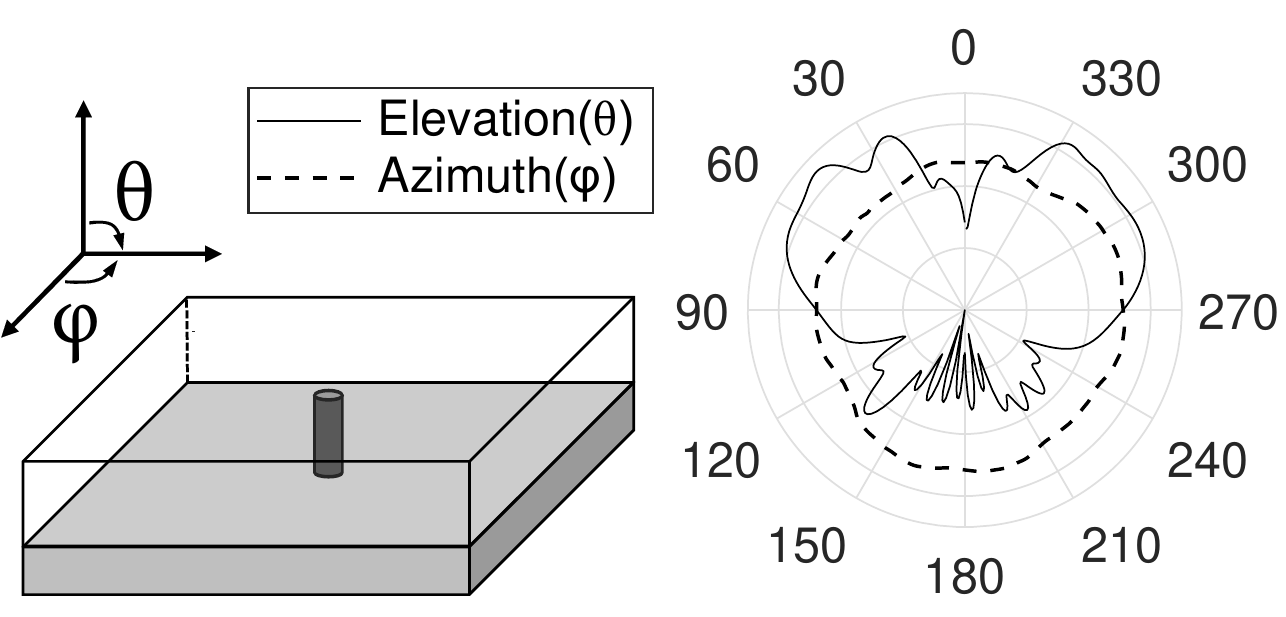}
\includegraphics[width=0.9\columnwidth]{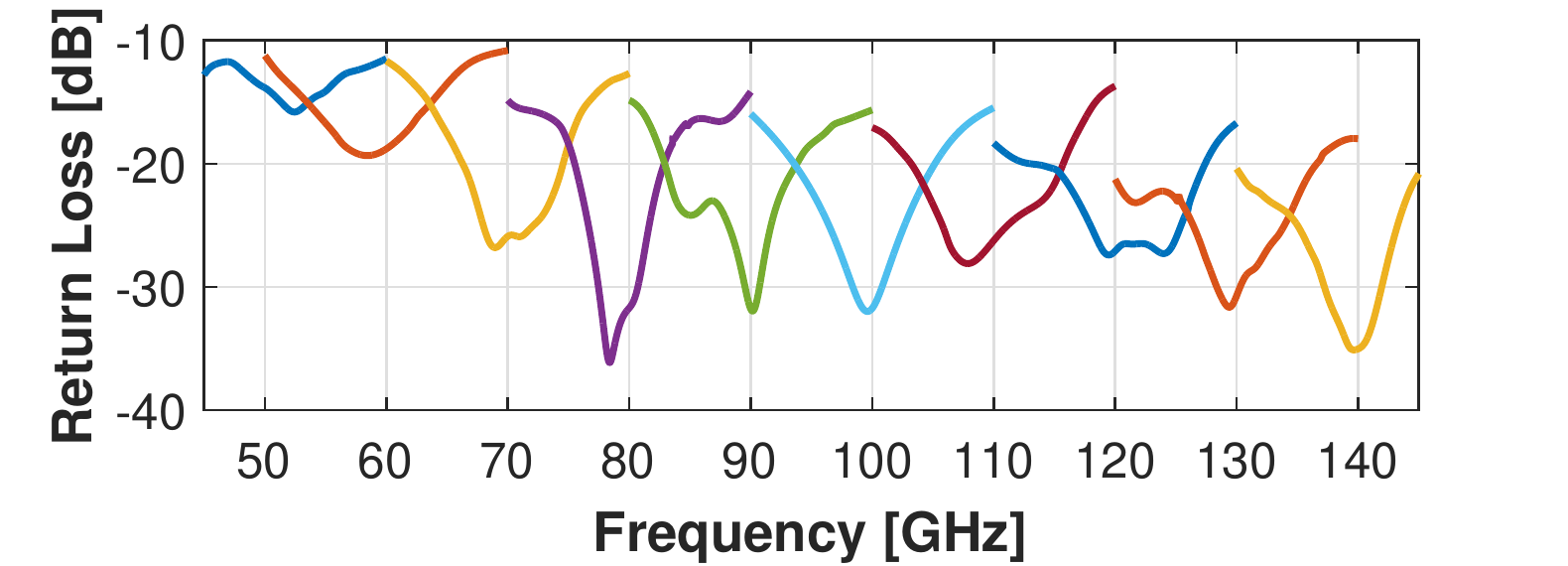}
\vspace{-0.2cm}
\caption{Schematic representation of an on-chip monopole, expected radiation pattern, and return loss for instantiations optimized at 50--140 GHz.}
\label{fig:monopole}\label{fig:s11freqsweep}
\end{figure}

The monopole antenna is modeled as a thin and long cylindrical metallic structure, placed vertically passing through the silicon. Through optimization-driven simulations, the length of the monopole is adjusted to minimize the return loss at the central frequency of interest. By default, simulations are by default centered at 60 GHz with 20 GHz bandwidth. However, explorations at higher frequencies are also performed after the respective monopole length adjustments. Figure \ref{fig:s11freqsweep} shows the return loss of the different monopole instances.

Simulations consider a number of antennas evenly distributed across the chips. The outcomes are the field distribution, the antenna gain, and the coupling between antennas. Let $\overline{S_{ij}}$ be the average of the coupling between transmitter $j$ and receiver $i$ over the whole frequency band. The minimum of $\overline{S}$ is used as a benchmark to evaluate the worst case for several material thickness combinations. It can be expressed as:
\vspace{-0.1cm}\begin{equation} \label{eq:Smin}
S_{min} = \min_{i,j\neq i}\overline{S_{ij}}.
\vspace{-0.1cm}\end{equation}
With the S-parameters, the channel frequency response $H_{ij}(f)$ can be then evaluated for each antenna pair as 
\begin{equation} \label{eq:Hf}
G_{i} G_{j} |H_{ij}(f)|^{2} = \frac{|S_{ji}(f)|^{2}}{(1 - |S_{ii}(f)|^{2})\cdot(1 - |S_{jj}(f)|^{2})},	
\end{equation}
where $G_{i}$ and $G_{j}$ are the transmitter and receiver antenna gains, $S_{ji}$ is the coupling between transmitter $i$ and receiver $j$, whereas $S_{ii}$ and $S_{jj}$ are the reflection coefficients at both ends \cite{Lin2007}. Once evaluated, a path loss analysis can be performed by fitting the attenuation over distance to
\begin{equation} \label{eq:2}
L_{dB} = 10n \cdot \log_{10}(d) + C,
\end{equation}
where $d$ is the distance between antennas and $n$ is the path loss exponent \cite{Zhang2007}. The path loss exponent is around 2 in free space, below 2 in guided or enclosed structures, and above 2 in lossy environments.

\section{Simulation Results}
\label{sec:simulationresults}
In the following, we show the results of an extensive simulation study that explores the channel characteristics in single-chip and multi-chip settings. We perform package optimizations to minimize path loss and assess the impact of having a multiple chips on the optimal design point. Additionally, we explore the scaling with frequency. 

\subsection{On-chip wireless channel in single-chip package}
We start by exploring the channel within a single-chip package. This models conventional processors, but also serves for 3D stacking if we assume that antennas are placed at the top chip --the one interfaced by the heat spreader. For this scenario, we consider a single square chip 22-mm long and wide, surrounded by conducting walls representing the package boundaries. The conducting walls are placed 5 mm away of the chip limits. The space between the chip and the walls is modeled as vacuum, although it could be filled with molding compound as discussed in Section \ref{sec:system-model}. 

\begin{figure*}[!t]
\centering
\subfigure[60 GHz\label{fig:gainopt60}]{\includegraphics[width=0.8\columnwidth]{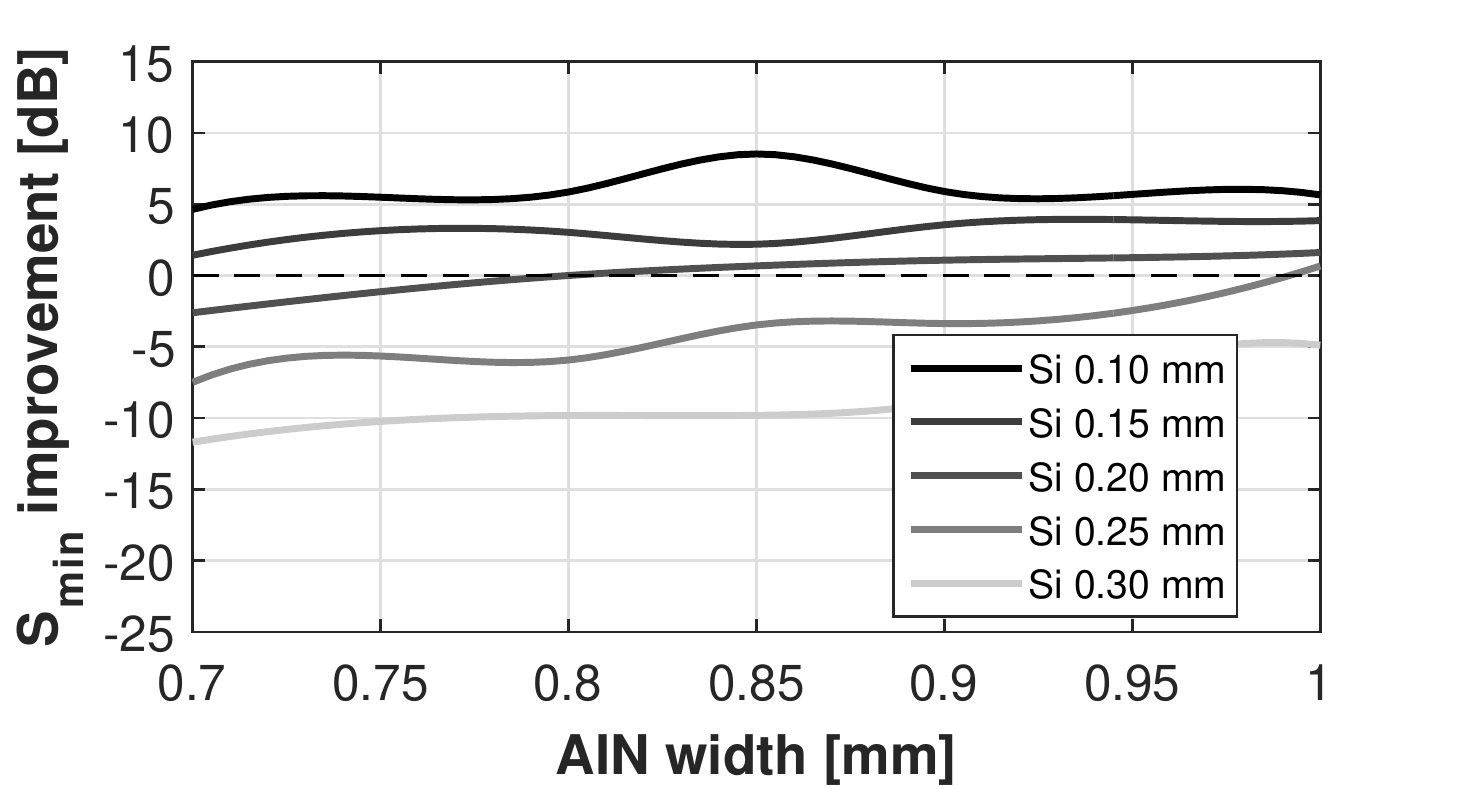}}
\subfigure[100 GHz\label{fig:gainopt100}]{\includegraphics[width=0.8\columnwidth]{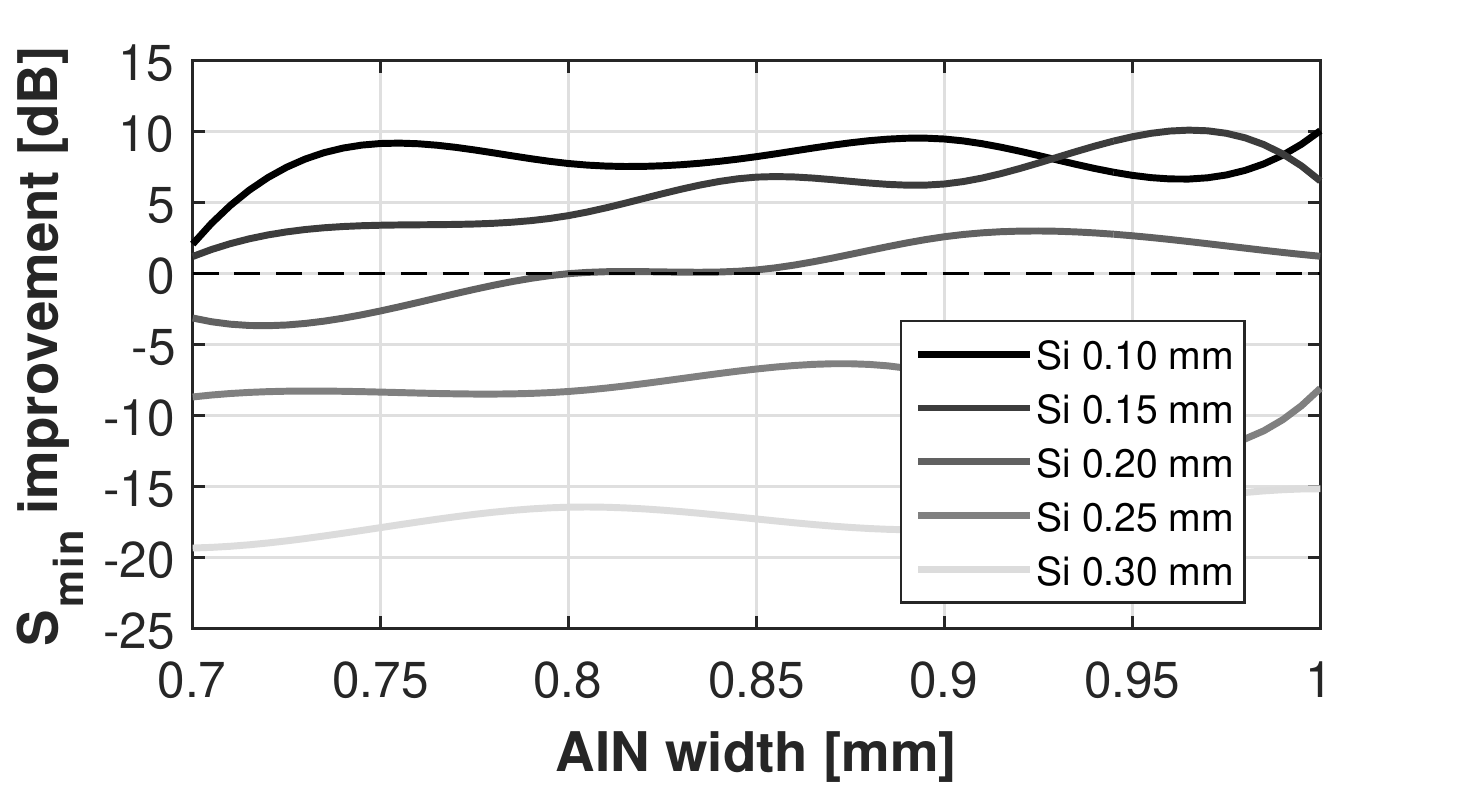}}
\vspace{-0.2cm}
\caption{Improvement of worst-case coupling $S_{min}$ in the single-chip case at two frequency points. The baseline dimensions are specified in Table \ref{tab:flipchip}.}
\label{fig:gainopt}
\end{figure*}

\textbf{Package Optimization at 60 GHz.} As indicated in Section \ref{sec:opti}, it is preferable to keep the silicon thickness to a minimum and to increase that of the heat spreader. When introducing the antenna, results may oscillate and an antenna-package co-design may be required for optimization. Fig. \ref{fig:gainopt60} shows the results of such co-design, which keeping the monopole matched at 60 GHz at all times. It is found that the optimal silicon and AIN thicknesses are 0.10 mm and 0.85 mm, respectively, as they yield the highest mean of the worst case coupling. This shows that fine-grained optimization can provide extra 5--10 dB of path loss reduction. 


\begin{figure}[!t]
\centering
\includegraphics[width=0.9\columnwidth]{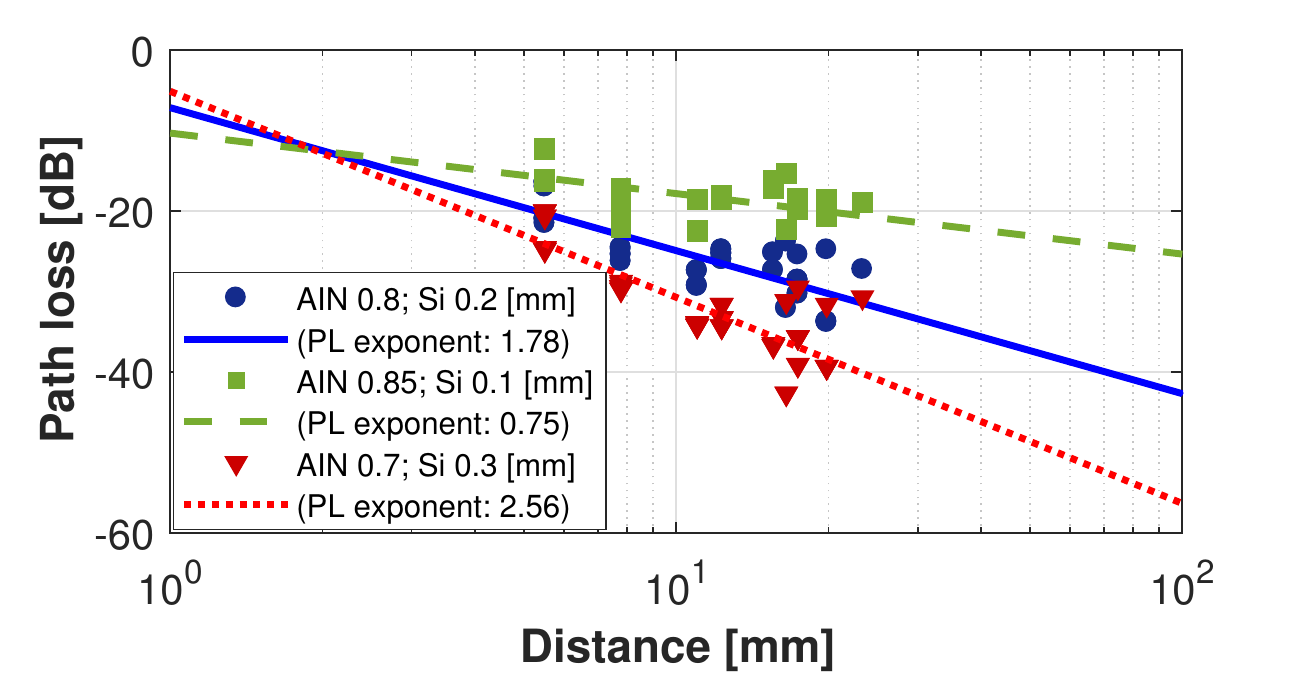}
\caption{Path loss as a function of distance, including linear regression fitting, for the single-chip case at 60 GHz.}
\label{fig:pathloss1chip22}
\end{figure}

\textbf{Path loss at 60 GHz.} To further highlight the importance of package optimization, we performed a path loss analysis at 60 GHz. We considered three different cases: default dimensions as specified in Table \ref{tab:flipchip}, optimal dimensions as obtained in Fig. \ref{fig:gainopt60}, and a quite suboptimal design point. Remind that the path loss decouples the antenna effects and leaves just losses due to propagation. The results, plotted in Figure \ref{fig:pathloss1chip22}, shows how optimization reduces not only the path loss overall, but also the path loss exponent. For the default case, the path loss exponent is 1.78, slightly lower than the free space path loss, thanks to having a confined environment. In the optimal case, we are able to cut the exponent down to 0.75, thereby showing a strong waveguiding effect in propagation. The suboptimal case, with an exponent higher than 2, demonstrates that the losses introduced by silicon cannot be neglected.


\begin{figure}[!t]
\centering
\includegraphics[width=\columnwidth]{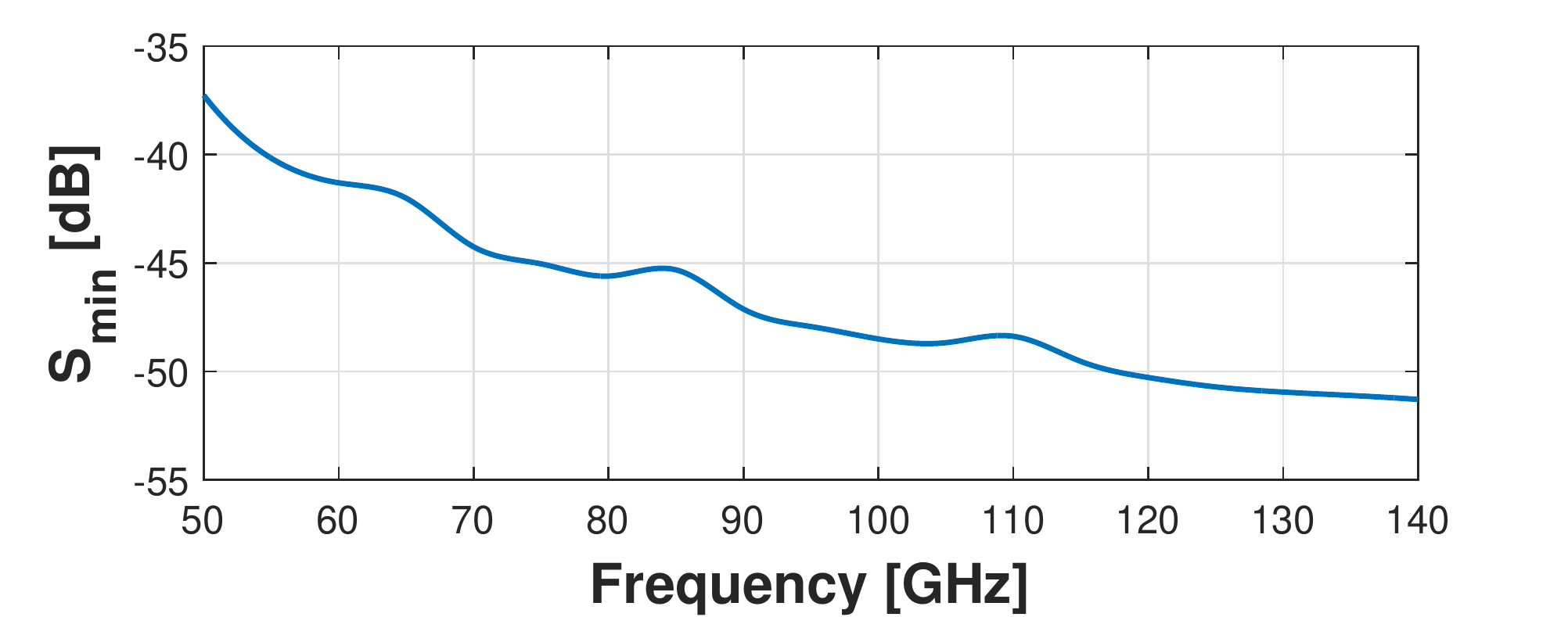}
\caption{Worst-case coupling $S_{min}$ as a function of frequency in the single-chip case with the thicknesses specified in Table \ref{tab:flipchip}.} 
\label{fig:gainvsfreq}
\end{figure}

\textbf{Frequency sweep.} Increasing the frequency of operation leads to smaller antennas and, potentially, smaller transceivers. Also, the absolute bandwidth is generally improved. Therefore, it is of great interest to study the scaling trends of the on-chip channel. Figure \ref{fig:gainvsfreq} shows how $S_{min}$ scales over frequency. This sweep was performed with the default silicon and heat spreader widths, 0.2 mm and 0.8 mm, respectively. We can observe that in overall, the loss between links increases with frequency probably due to two reasons: the antennas have smaller apertures, and the propagation losses at the dielectrics are larger. Nevertheless, this effect is compensated in part by the enclosed nature of the on-chip scenario, mitigating the impacts of frequency scaling. 

\textbf{Impact of frequency on package optimization.} Since our methodology performs a joint antenna-package optimization, it is reasonable to think that the optimal point will change with frequency. To illustrate this, we performed the exploration with the monopole tuned at 100 GHz. Figure \ref{fig:gainopt100} shows how the optimal point has slightly changed, but the tendency of higher losses with a thicker silicon is indeed, increased. This can be explained by the fact that losses on the silicon are frequency sensitive. Still, the improvement with respect to the default case is 10 dB and can be achieved even with 0.15mm of silicon. 

\subsection{On-chip wireless channel in multi-chip package}
Let us now consider a single chip isolated in a package without lateral walls. This would \emph{a priori} model the worst case for on-chip propagation in a multi-chip package, where lateral walls are far away and neighboring chips absorb most of the incoming energy. Indeed, the absence of reflecting elements nearby is expected to lead to a reduction of the energy that returns to the chip after leaving, therefore increasing the path loss. To evaluate this scenario, we consider a 22$\times$22 mm\textsuperscript{2} chip whose boundaries model a perfect matching layer (PML), giving the impression of an infinitely wide package without walls. 

\begin{figure*}[!t]
\centering
\subfigure[60 GHz\label{fig:gainvsain60nowalls}]{\includegraphics[width=0.8\columnwidth]{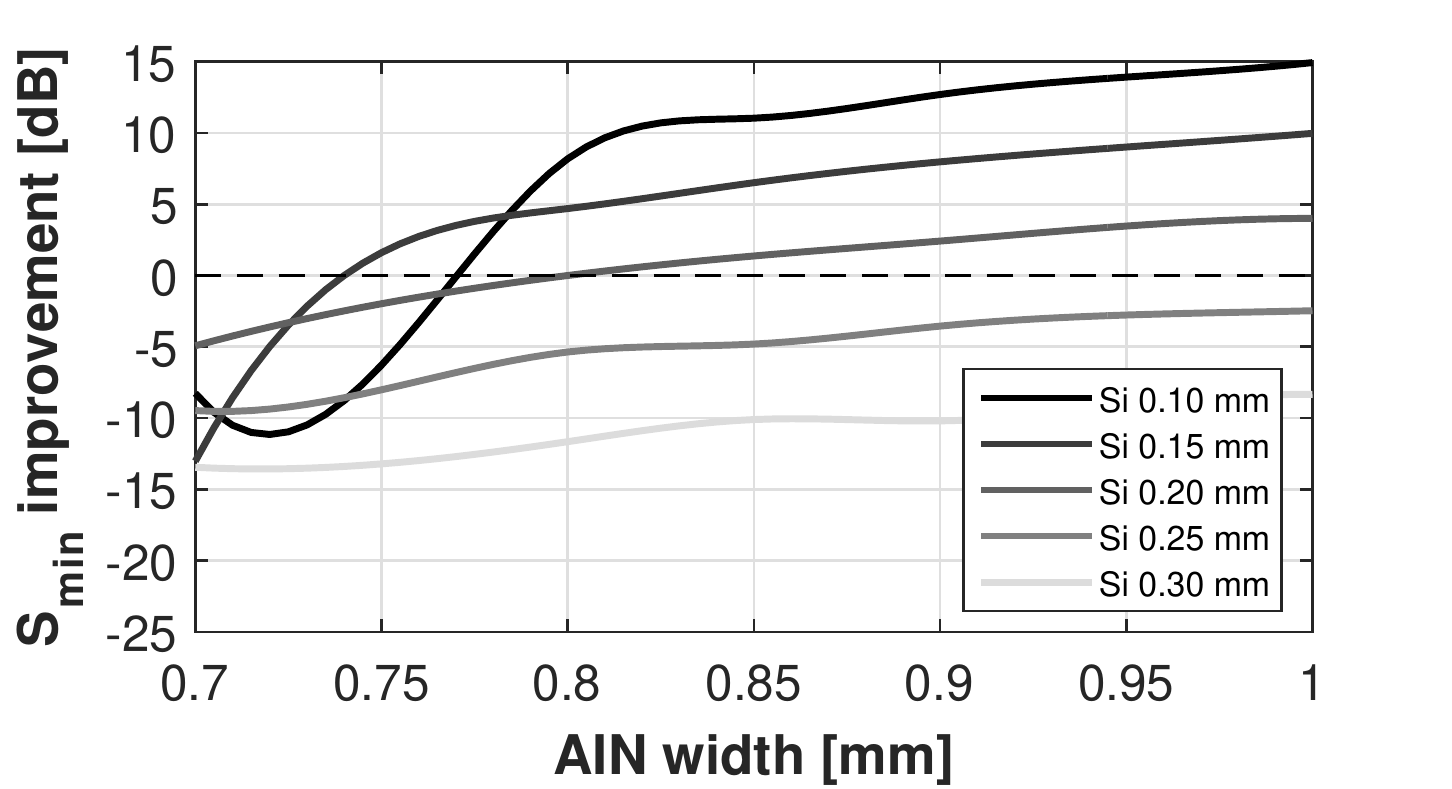}}
\subfigure[100 GHz\label{fig:gainvsain100nowalls}]{\includegraphics[width=0.8\columnwidth]{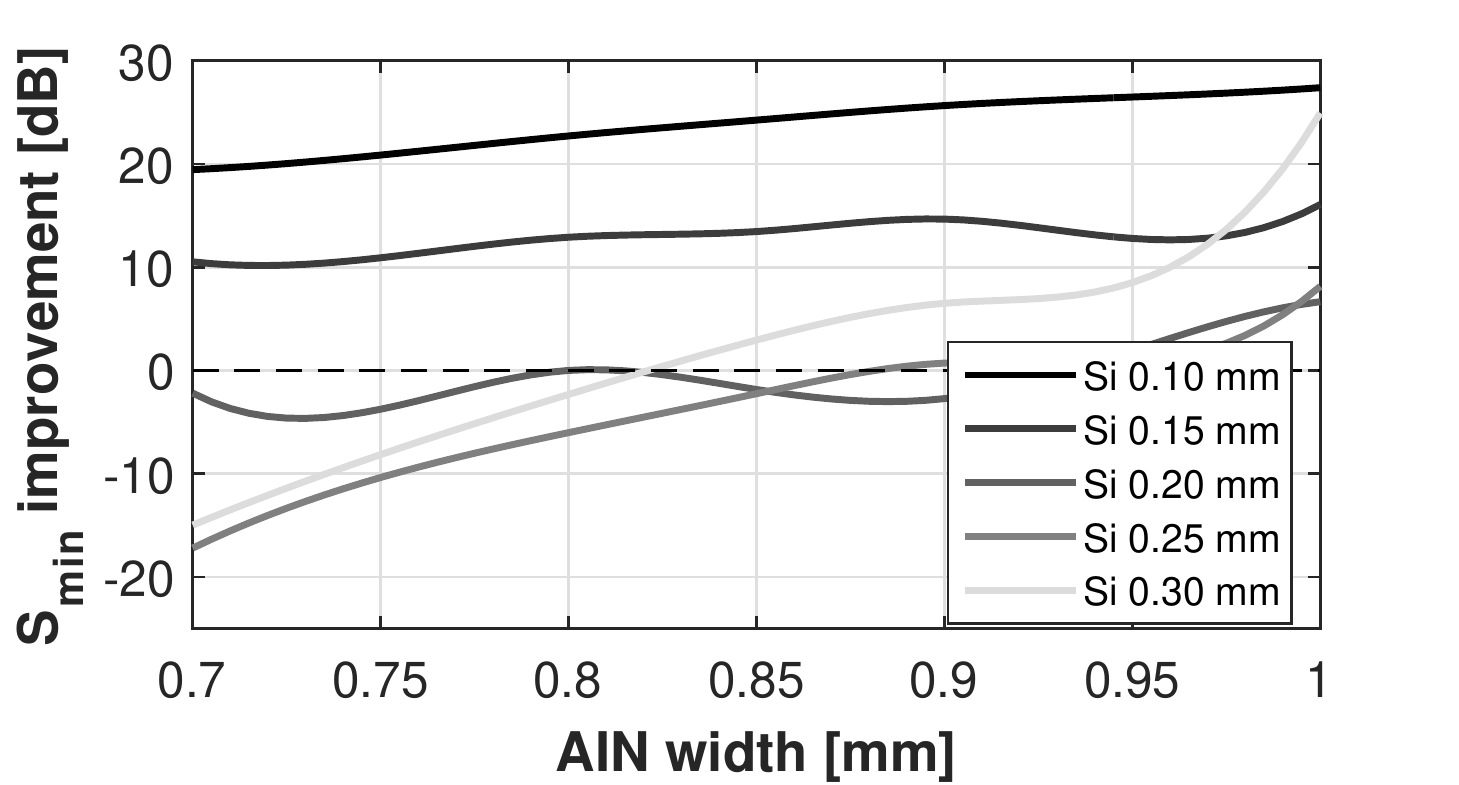}}
\vspace{-0.2cm}
\caption{Improvement of worst-case coupling $S_{min}$ in the multi-chip case (without lateral walls) at two frequency points. The baseline dimensions are specified in Table \ref{tab:flipchip}.}	
\end{figure*} 



\textbf{Impact of boundaries on loss between links.} Without lateral walls, the loss of the worst-case link is increased up to 27 dB higher for the thicknesses of Table \ref{tab:flipchip} (from -43 dB to -70 dB). This is due to the harsh reduction of the power received by the antenna at the opposite corner of the radiating one, as the majority of the power to this antenna was coming from the now non-existent reflections at the package walls. This results strongly suggests that, being either due to losses or security reasons (so that wireless signals cannot escape outside the package), having package walls is beneficial. 

\textbf{Impact of boundaries on package optimization.} As we can see on Figures \ref{fig:gainvsain60nowalls} and \ref{fig:gainvsain100nowalls}, the variation of $S_{min}$ with respect to non-optimized case is large and more sensitive to the silicon thickness. Due to the absence of walls, package optimization plays an even more important role than with walls (Fig. \ref{fig:gainopt}). The improvement is larger than 20 dB in several cases. Note, also, that the exploration also unveiled very detrimental $S_{min}$ dips at certain thickness combinations. See, for instance, the 0.1-mm silicon and 0.7-mm AIN case at 60 GHz in Figure \ref{fig:gainvsain60nowalls}. 

\subsection{Off-chip wireless channel in multi-chip package}
We finally consider a full multi-chip packages as represented in Figures \ref{fig:interposer} and \ref{fig:MCM}. In the interposer case, we evaluate an array of 2$\times$2 small chips, 10 mm in length each, placed inside a 33$\times$33 mm\textsuperscript{2} package with a separation of 5 mm between chips. The silicon interposer has 0.1 mm of thickness and 33 mm of length and width. In the MCM case, we simulate 2$\times$2 chips of standard size, placed inside a bigger package whose length and width is 59 mm. 

We place four antennas per chip, regardless of the chip size, in order to evaluate all the possible combinations, i.e., close or distant antennas in close or distant chips. All the multichip package simulations are performed with enclosing conducting walls and a common heat spreader for all chips.

\begin{figure*}[!t]
\centering
\subfigure[Single standard chip on package substrate.\label{fig:min22x1}]{\includegraphics[width=0.8\columnwidth]{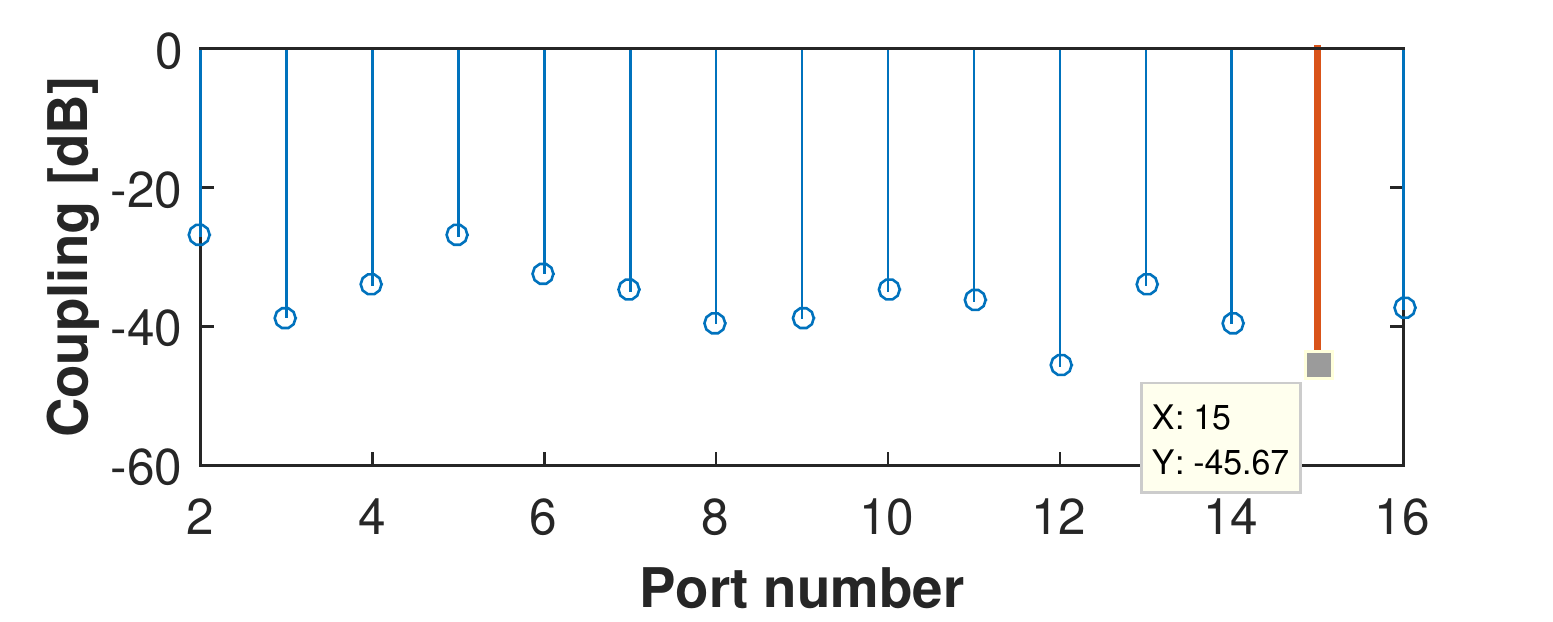}}
\subfigure[Four small chips on silicon interposer.\label{fig:min10x4}]{\includegraphics[width=0.8\columnwidth]{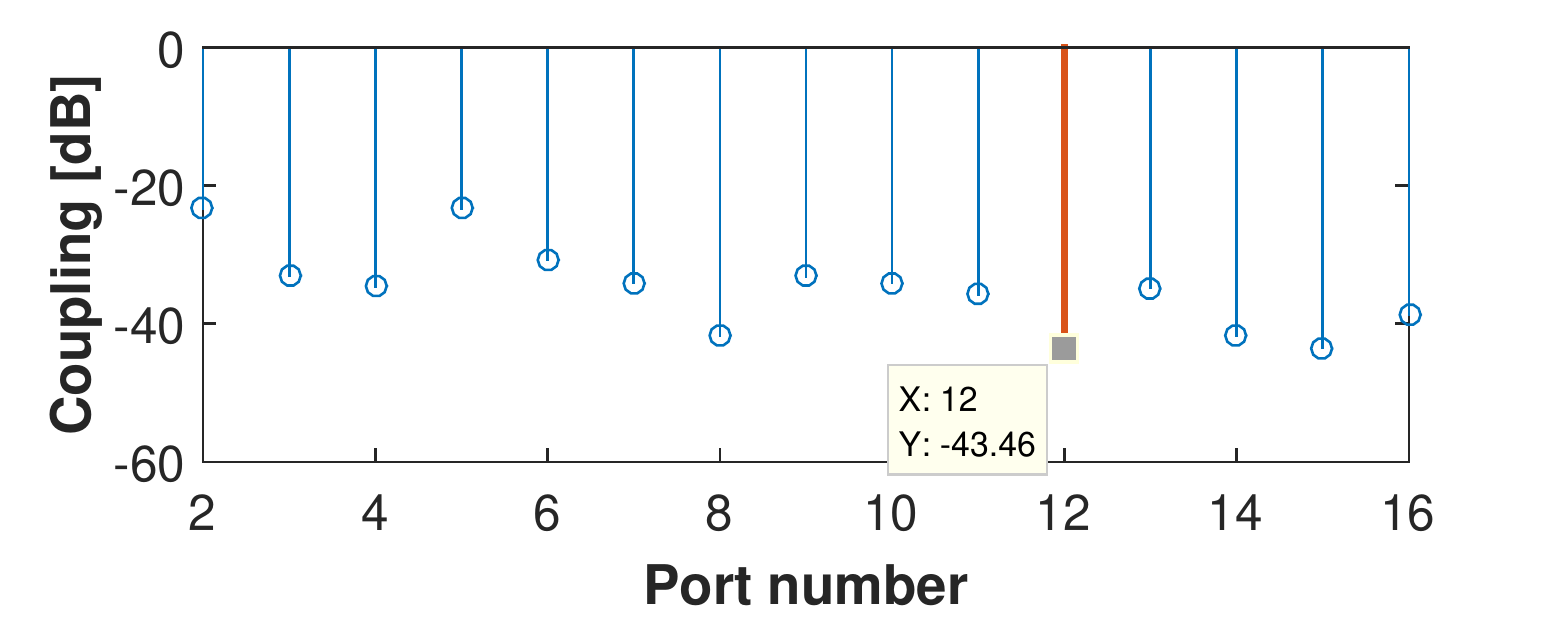}}
\vspace{-0.2cm}
\caption{Mean of the S-parameter for each antenna pair assuming transmission from port 1 at 60 GHz. $S_{min}$ is highlighted.}
\label{fig:minmeanscomparison}
\end{figure*}

\textbf{Small chips vs single standard chip.} The silicon interposer case allows us to evaluate the impact of processor disintegration \cite{Kannan2016} on the wireless channel characteristics. To this end, we compare the coupling between antennas in a single large chip, Fig. \ref{fig:min22x1}, and multiple small chips, Fig. \ref{fig:min10x4}. The plots do not illustrate large changes overall --a slightly better coupling is observed in the interposer case (only a few dBs in most antenna pairs, including $S_{min}$). There are two effects that seem to be canceling out: on the one hand, propagation occurring in the vacuum space between chips instead of in lossy silicon would lead to better coupling in the interposer case. On the other hand, reflections due to media changes (silicon--vacuum--silicon) are also higher in the interposer case, which may be leading to lower coupling in far away antennas. This may also explain the better coupling at nearby antennas (port 2 and 5).


\begin{figure*}[!t]
\centering
\subfigure[Four small chips on silicon interposer.\label{fig:pathloss4chip10}]{\includegraphics[width=0.8\columnwidth]{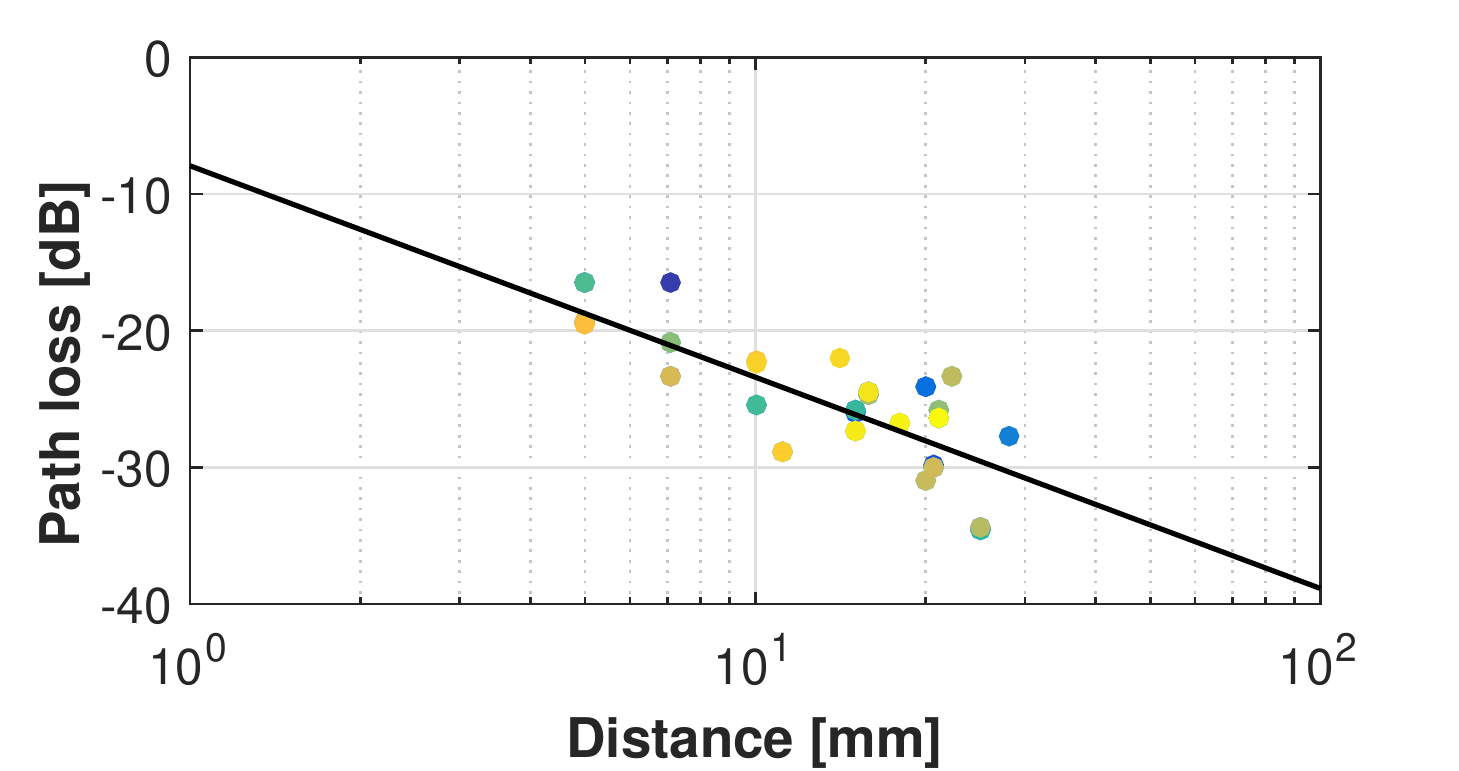}}
\subfigure[Four standard chips on MCM package.\label{fig:pathloss4chip22}]{\includegraphics[width=0.8\columnwidth]{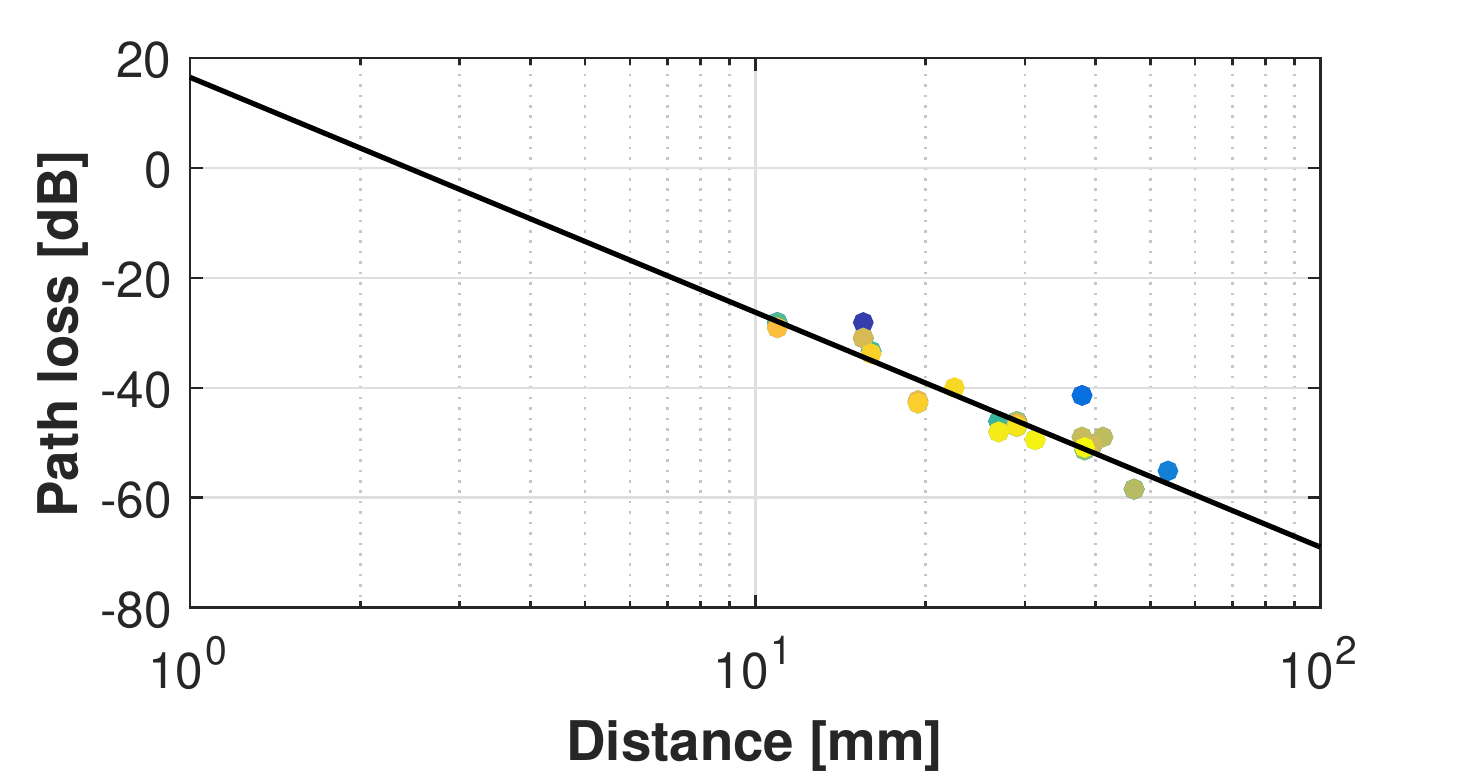}}
\vspace{-0.2cm}
\caption{Path loss as a function of distance, including linear regression fitting, for the multi-chip cases at 60 GHz.}
\label{fig:pathloss4chip}
\end{figure*}

\textbf{Path loss: interposer vs MCM.} The path loss exponent is calculated for the two multi-chip scenarios and compared with that of 3D stacking, previously shown in Fig. \ref{fig:pathloss1chip22}. In all cases, we considered the default thicknesses specified in Table \ref{tab:flipchip}. The path loss exponent of the interposer scenario, plotted in Figure \ref{fig:pathloss4chip10}, is 1.55. This is slightly lower than the exponent of the single-chip case (1.78), thanks to the lack of silicon between chips. The path loss exponent increases up to 4.27 for MCM case, plotted in Fig. \ref{fig:pathloss4chip22}, due to the crescent losses due to propagation through lossy silicon as the distance between antennas increases. This confirms that the amount of silicon that waves need to traverse is the main determinant of losses.

\section{Related Work}
\label{sec:related}		
\textbf{Channels within a Chip Package:} the existence of a wireless channel in flip-chip packages was suggested in \cite{Kim2001} and then experimentally validated at 15 GHz in \cite{Branch2005}. Recently, the work by Narde \emph{et al.} analyzes the S-parameters for one and two chips within the same package assuming planar zig-zag antennas at 60 GHz \cite{Narde2018}. Although becoming part of the radiative structure by capacitive coupling, the effect of bumps on the antenna response is not discussed. They also assume high-resistivity silicon, which may be unlikely in processor dies. Our previous works, instead, discuss different 60-GHz antenna types and perform field distribution and path loss analysis within a single-chip package with bulk silicon \cite{Timoneda2018, Timoneda2018a}. The present work extends the frequency range up to 150 GHz and considers multiple single-chip and multi-chip configurations, both \ac{MCM} and interposer-based.

\textbf{Channels within a Metallic Casing:} Matolak \emph{et al.} suggested that the wireless chip-scale communications would act as a micro-reverberation chamber with metallic walls \cite{Matolak2013CHANNEL}, although they did not discuss the package. Others have explored a similar scenario over a large PCB board, including DRAM modules and other components within a computer case up to 300 GHz \cite{Chen2007prop, Chiang2010, Wu2013a, Kim2016mother}. In \cite{Tasolamprou2018}, the authors explored waveguide-like millimeter-wave channels within a reconfigurable metamaterial that integrates multiple chips on a PCB. Finally, the 60-GHz channel has been also studied in larger enclosed environments such as printers \cite{Ohira2011} or data center cabinets \cite{Khademi2015}, which also act as reverberation chambers. Although these structures may capture the enclosed nature of a chip package, they have substantially different dimensions, materials, and antenna placement restrictions. 

\textbf{Chip-scale Channels without Package:} studies of the on-chip and off-chip wireless channels have been often conducted without considering any particular package, assuming free space over the insulator layer. Yan \emph{et al.} provided a theoretical basis at millimeter-wave frequencies \cite{Yan2009}, whereas others provided simulation-based studies \cite{Liu2015, Narde2017, Gade2017} or actual measurements using planar antennas \cite{Zhang2007, Yeh2013} and bond-wire antennas \cite{Chen2009}. In the terahertz band analysis of \cite{Chen2018}, the package structure is described, but then neglected for simplicity. Finally, monopoles placed in a loosely defined superstrate have been also studied in recent works \cite{Rayess2017, Wu2017b}. All these works, however, do not provide a faithful view of a chip package and cannot be re-used for the scenario at hand.

\section{Conclusion}
\label{sec:conclusion}
The characterization of the wireless channel in scenarios compatible with standard chip packages is largely missing in the literature. To start bridging this gap, here we have performed a frequency-domain analysis of mm-wave propagation in the 3D stacking, silicon interposer, and multi-chip module schemes. We highlight the importance of package optimization to ensure the feasibility of the WNoC approach, as it is capable of reducing path losses by several tens of dBs. We also conclude that such optimization is dependent on the frequency of operation	 and the elements surrounding the chips. Finally, a path loss analysis confirms that propagation length within silicon is the main determinant of losses, and that finding the right package dimensions ensure the scalability of the approach.






\begin{thebibliography}{10}
\providecommand{\url}[1]{#1}
\csname url@samestyle\endcsname
\providecommand{\newblock}{\relax}
\providecommand{\bibinfo}[2]{#2}
\providecommand{\BIBentrySTDinterwordspacing}{\spaceskip=0pt\relax}
\providecommand{\BIBentryALTinterwordstretchfactor}{4}
\providecommand{\BIBentryALTinterwordspacing}{\spaceskip=\fontdimen2\font plus
\BIBentryALTinterwordstretchfactor\fontdimen3\font minus
  \fontdimen4\font\relax}
\providecommand{\BIBforeignlanguage}[2]{{%
\expandafter\ifx\csname l@#1\endcsname\relax
\typeout{** WARNING: IEEEtran.bst: No hyphenation pattern has been}%
\typeout{** loaded for the language `#1'. Using the pattern for}%
\typeout{** the default language instead.}%
\else
\language=\csname l@#1\endcsname
\fi
#2}}
\providecommand{\BIBdecl}{\relax}
\BIBdecl

\bibitem{Arunkumar2017}
A.~Arunkumar, E.~Bolotin, B.~Cho, U.~Milic, E.~Ebrahimi, O.~Villa, A.~Jaleel,
  C.-J. Wu, and D.~Nellans, ``{MCM-GPU: Multi-Chip-Module GPUs for Continued
  Performance Scalability},'' \emph{Proceedings of the ISCA '17}, pp. 320--332,
  2017.

\bibitem{Kannan2016}
A.~Kannan, N.~{Enright Jerger}, and G.~H. Loh, ``{Exploiting Interposer
  Technologies to Disintegrate and Reintegrate Multicore Processors},''
  \emph{IEEE Micro}, vol.~36, no.~3, pp. 84--93, 2016.

\bibitem{Yin2018}
J.~Yin, Z.~Lin, O.~Kayiran, M.~Poremba, M.~S. {Bin Altaf}, N.~{Enright Jerger},
  and G.~H. Loh, ``{Modular Routing Design for Chiplet-based Systems},''
  \emph{Proceedings of the ISCA '18}, 2018.

\bibitem{Zhang2015d}
X.~Zhang, J.~K. Lin, S.~Wickramanayaka, S.~Zhang, R.~Weerasekera, R.~Dutta,
  K.~F. Chang, K.~J. Chui, H.~Y. Li, D.~S. {Wee Ho}, L.~Ding, G.~Katti,
  S.~Bhattacharya, and D.~L. Kwong, ``{Heterogeneous 2.5D integration on
  through silicon interposer},'' \emph{Applied Physics Reviews}, vol.~2, no.~2,
  2015.

\bibitem{Kim2012Survey}
J.~Kim, K.~Choi, and G.~Loh, ``{Exploiting new interconnect technologies in
  on-chip communication},'' \emph{IEEE Journal on Emerging and Selected Topics
  in Circuits and Systems}, vol.~2, no.~2, pp. 124--136, 2012.

\bibitem{Thraskias2018}
C.~Thraskias, E.~Lallas, N.~Neumann, L.~Schares, B.~Offrein, R.~Henker,
  D.~Plettemeier, F.~Ellinger, J.~Leuthold, and I.~Tomkos, ``{Survey of
  Photonic and Plasmonic Interconnect Technologies for Intra-Datacenter and
  High-Performance Computing Communications},'' \emph{IEEE Communications
  Surveys and Tutorials}, 2018.

\bibitem{Sujay2012}
S.~Deb, A.~Ganguly, P.~P. Pande, B.~Belzer, and D.~Heo, ``{Wireless NoC as
  Interconnection Backbone for Multicore Chips: Promises and Challenges},''
  \emph{IEEE Journal on Emerging and Selected Topics in Circuits and Systems},
  vol.~2, no.~2, pp. 228--239, 2012.

\bibitem{Chen2009}
W.-H. Chen, S.~Joo, S.~Sayilir, R.~Willmot, T.-Y. Choi, D.~Kim, J.~Lu,
  D.~Peroulis, and B.~Jung, ``{A 6-Gb/s Wireless Inter-Chip Data Link Using
  43-GHz Transceivers and Bond-Wire Antennas},'' \emph{IEEE Journal of
  Solid-State Circuits}, vol.~44, no.~10, pp. 2711--2721, oct 2009.

\bibitem{Fettweis2013}
G.~P. Fettweis, N.~ul~Hassan, L.~Landau, and E.~Fischer, ``{Wireless
  Interconnect for Board and Chip Level},'' in \emph{Proceedings of the DATE
  '13}.\hskip 1em plus 0.5em minus 0.4em\relax New Jersey: IEEE Conference
  Publications, 2013, pp. 958--963.

\bibitem{Baniya2018a}
P.~Baniya, S.~Yoo, K.~L. Melde, A.~Bisognin, and C.~Luxey, ``{Switched-Beam
  60-GHz Four-Element Array for Multichip Multicore System},'' \emph{IEEE
  Transactions on Components, Packaging and Manufacturing Technology}, vol.~8,
  no.~2, pp. 251--260, 2018.

\bibitem{Yu2014}
X.~Yu, J.~Baylon, P.~Wettin, D.~Heo, P.~{Pratim Pande}, and S.~Mirabbasi,
  ``{Architecture and Design of Multi-Channel Millimeter-Wave Wireless
  Network-on-Chip},'' \emph{IEEE Design {\&} Test}, vol.~31, no.~6, pp. 19--28,
  2014.

\bibitem{Palesi2015}
M.~Palesi, M.~Collotta, A.~Mineo, and V.~Catania, ``{An Efficient Radio Access
  Control Mechanism for Wireless Network-On-Chip Architectures},''
  \emph{Journal of Low Power Electronics and Applications}, vol.~5, no.~2, pp.
  38--56, 2015.

\bibitem{Abadal2018a}
S.~Abadal, A.~Mestres, J.~Torrellas, E.~Alarc{\'{o}}n, and
  A.~Cabellos-Aparicio, ``{Medium Access Control in Wireless Network-on-Chip: A
  Context Analysis},'' \emph{IEEE Communications Magazine}, vol.~56, no.~6, pp.
  172--178, 2018.

\bibitem{DiTomaso2015}
D.~DiTomaso, A.~Kodi, D.~Matolak, S.~Kaya, S.~Laha, and W.~Rayess, ``{A-WiNoC:
  Adaptive Wireless Network-on-Chip Architecture for Chip Multiprocessors},''
  \emph{IEEE Transactions on Parallel and Distributed Systems}, vol.~26,
  no.~12, pp. 3289--3302, 2015.

\bibitem{Abadal2018}
S.~Abadal, J.~Torrellas, E.~Alarc{\'{o}}n, and A.~Cabellos-Aparicio,
  ``{OrthoNoC: A Broadcast-Oriented Dual-Plane Wireless Network-on-Chip
  Architecture},'' \emph{IEEE Transactions on Parallel and Distributed
  Systems}, vol.~29, no.~3, pp. 628--641, 2018.

\bibitem{AbadalASPLOS}
S.~Abadal, E.~Alarc{\'{o}}n, A.~Cabellos-Aparicio, and J.~Torrellas, ``{WiSync:
  An Architecture for Fast Synchronization through On-Chip Wireless
  Communication},'' in \emph{Proceedings of the ASPLOS '16}, 2016, pp. 3--17.

\bibitem{Choi2018}
W.~Choi, K.~Duraisamy, R.~G. Kim, J.~R. Doppa, P.~P. Pande, D.~Marculescu, and
  R.~Marculescu, ``{On-Chip Communication Network for Efficient Training of
  Deep Convolutional Networks on Heterogeneous Manycore Systems},'' \emph{IEEE
  Transactions on Computers}, vol.~67, no.~5, pp. 672--686, 2018.

\bibitem{Gade2017a}
S.~H. Gade and S.~Deb, ``{HyWin: Hybrid wireless NoC with sandboxed
  sub-networks for CPU/GPU architectures},'' \emph{IEEE Transactions on
  Computers}, vol.~66, no.~7, pp. 1145--1158, 2017.

\bibitem{Sikder2016}
M.~A.~I. Sikder, A.~Kodi, W.~Rayess, D.~Ditomaso, D.~Matolak, and S.~Kaya,
  ``{Exploring wireless technology for off-chip memory access},'' in
  \emph{Proceedings of the HOTI '16}, 2016, pp. 92--99.

\bibitem{Liu2016}
Z.~Liu, Y.~Liang, N.~Li, G.~Feng, H.~Yu, and S.~Chen, ``{An Energy-efficient
  Adaptive Sub-THz Wireless Interconnect with MIMO-Beamforming between Cores
  and DRAMs},'' in \emph{Proceedings of the NANOCOM '16}, 2016, pp. 1--6.

\bibitem{Shamim2017}
S.~Shamim, N.~Mansoor, R.~S. Narde, V.~Kothandapani, A.~Ganguly, and
  J.~Venkataraman, ``{A Wireless Interconnection Framework for Seamless Inter
  and Intra-chip Communication in Multichip Systems},'' \emph{IEEE Transactions
  on Computers}, vol.~66, no.~3, pp. 389--402, 2017.

\bibitem{Matolak2013CHANNEL}
D.~Matolak, S.~Kaya, and A.~Kodi, ``{Channel modeling for wireless
  networks-on-chips},'' \emph{IEEE Communications Magazine}, vol.~51, no.~6,
  pp. 180--186, 2013.

\bibitem{Zhang2007}
Y.~P. Zhang, Z.~M. Chen, and M.~Sun, ``{Propagation Mechanisms of Radio Waves
  Over Intra-Chip Channels With Integrated Antennas: Frequency-Domain
  Measurements and Time-Domain Analysis},'' \emph{IEEE Transactions on Antennas
  and Propagation}, vol.~55, no.~10, pp. 2900--2906, 2007.

\bibitem{Rayess2017}
W.~Rayess, D.~W. Matolak, S.~Kaya, and A.~K. Kodi, ``{Antennas and Channel
  Characteristics for Wireless Networks on Chips},'' \emph{Wireless Personal
  Communications}, vol.~95, no.~4, pp. 5039--5056, 2017.

\bibitem{Chen2018}
Y.~Chen and C.~Han, ``{Channel Modeling and Analysis for Wireless
  Networks-on-Chip Communications in the Millimeter Wave and Terahertz
  Bands},'' in \emph{Proceedings of the INFOCOM WKSHPS '18}, 2018.

\bibitem{Chen2007prop}
Z.~Chen and Y.~Zhang, ``{Inter-chip wireless communication channel:
  Measurement, characterization, and modeling},'' \emph{IEEE Transactions on
  Antennas and Propagation}, vol.~55, no.~3, pp. 978--986, 2007.

\bibitem{Narde2017}
R.~S. Narde and J.~Venkataraman, ``{Feasibility study of Transmission between
  Wireless Interconnects in Multichip Multicore systems},'' in
  \emph{Proceedings of the APS/URSI '17}, 2017, pp. 1821--1822.

\bibitem{Baniya2018}
P.~Baniya, A.~Bisognin, K.~L. Melde, and C.~Luxey, ``{Chip-to-Chip Switched
  Beam 60 GHz Circular Patch Planar Antenna Array and Pattern
  Considerations},'' \emph{IEEE Transactions on Antennas and Propagation},
  vol.~66, no.~4, pp. 1776--1787, 2018.

\bibitem{Chiang2010}
P.~Y. Chiang, S.~Woracheewan, C.~Hu, L.~Guo, H.~Liu, R.~Khanna, and J.~Nejedlo,
  ``{Short-Range, Wireless Interconnect within a Computing Chassis: Design
  Challenges},'' \emph{IEEE Design {\&} Test of Computers}, vol.~27, no.~4, pp.
  32--43, 2010.

\bibitem{Wu2013a}
H.-t. Wu, J.-j. Lin, and K.~K. O, ``{Inter-Chip Wireless Communication},'' in
  \emph{Proceedings of the EuCAP '13}, 2013, pp. 3647--3649.

\bibitem{Kim2016mother}
S.~Kim and A.~Zajic, ``{Characterization of 300 GHz Wireless Channel on a
  Computer Motherboard},'' \emph{IEEE Transactions on Antennas and
  Propagation}, vol.~64, no.~12, pp. 5411--5423, 2016.

\bibitem{Kim2001}
K.~Kim, W.~Bornstad, and K.~K. O, ``{A Plane Wave Model Approach to
  Understanding Propagation in an Intra-chip Communication System},'' in
  \emph{Proceedings of the APS '01}, 2001, pp. 166--169.

\bibitem{Branch2005}
J.~Branch, X.~Guo, L.~Gao, A.~Sugavanam, J.~J. Lin, and K.~K. O, ``{Wireless
  communication in a flip-chip package using integrated antennas on silicon
  substrates},'' \emph{IEEE Electron Device Letters}, vol.~26, no.~2, pp.
  115--117, 2005.

\bibitem{Narde2018}
R.~S. Narde, N.~Mansoor, A.~Ganguly, and J.~Venkataraman, ``{On-Chip Antennas
  for Inter-Chip Wireless Interconnections: Challenges and Opportunities},'' in
  \emph{Proceedings of the EuCAP '18}, 2018.

\bibitem{Topol2006}
A.~W. Topol, D.~C. {La Tulipe}, L.~Shi, D.~J. Frank, K.~Bernstein, S.~E. Steen,
  A.~Kumar, G.~U. Singco, A.~M. Young, K.~W. Guarini, and M.~Ieong,
  ``{Three-dimensional integrated circuits},'' \emph{IBM Journal of Research
  and Development}, vol.~50, no.~4, pp. 491--506, 2006.

\bibitem{ardebili2009encapsulation}
H.~Ardebili and M.~Pecht, \emph{Encapsulation technologies for electronic
  applications}.\hskip 1em plus 0.5em minus 0.4em\relax William Andrew, 2009.

\bibitem{Pal2018}
S.~Pal, D.~Petrisko, A.~A. Bajwa, P.~Gupta, S.~S. Iyer, and R.~Kumar, ``{A Case
  for Packageless Processors},'' in \emph{Proceedings of the HPCA-24}, 2018,
  pp. 466--479.

\bibitem{Timoneda2018}
X.~Timoneda, S.~Abadal, A.~Cabellos-Aparicio, D.~Manessis, J.~Zhou,
  A.~Franques, J.~Torrellas, and E.~Alarc{\'{o}}n, ``{Millimeter-Wave
  Propagation within a Computer Chip Package},'' in \emph{Proceedings of the
  ISCAS '18}, 2018.

\bibitem{Wright2006}
S.~L. Wright, R.~Polastre, H.~Gan, L.~P. Buchwalter, R.~Horton, P.~S. Andry,
  E.~Sprogis, C.~Patel, C.~Tsang, J.~Knickerbocker, J.~R. Lloyd, A.~Sharma, and
  M.~S. Sri-Jayantha, ``{Characterization of micro-bump C4 interconnects for
  Si-carrier SOP applications},'' in \emph{Proceedings of the ECTC '06}, 2006,
  pp. 633--640.

\bibitem{Kimoto2009}
K.~Kimoto, N.~Sasaki, S.~Kubota, W.~Moriyama, and T.~Kikkawa, ``{High-Gain
  On-Chip Antennas for LSI Intra- / Inter-Chip Wireless Interconnection},''
  \emph{Proceedings of the EuCAP '09}, pp. 278--282, 2009.

\bibitem{Markish2015}
O.~Markish, B.~Sheinman, O.~Katz, D.~Corcos, and D.~Elad, ``{On-chip mmWave
  Antennas and Transceivers},'' in \emph{Proceedings of the NoCS '15}, 2015, p.
  Art. 11.

\bibitem{Bieck2010}
F.~Bieck, S.~Spiller, F.~Molina, M.~T{\"{o}}pper, C.~Lopper, I.~Kuna, T.~C.
  Seng, and T.~Tabuchi, ``{Carrierless design for handling and processing of
  ultrathin wafers},'' \emph{Proceedings of the ECTC '10}, pp. 316--322, 2010.

\bibitem{Zhang2005}
Y.~P. Zhang, M.~Sun, and L.~H. Guo, ``{On-chip antennas for 60-GHz radios in
  silicon technology},'' \emph{IEEE Transactions on Electron Devices}, vol.~52,
  no.~7, pp. 1664--1668, 2005.

\bibitem{Gutierrez2009}
F.~Gutierrez, S.~Agarwal, K.~Parrish, and T.~S. Rappaport, ``{On-chip
  integrated antenna structures in CMOS for 60 GHz WPAN systems},'' \emph{IEEE
  Journal on Selected Areas in Communications}, vol.~27, no.~8, pp. 1367--1378,
  2009.

\bibitem{Yordanov2016}
H.~Yordanov, V.~Poulkov, and P.~Russer, ``{On-Chip Monolithic Integrated
  Antennas Using CMOS Ground Supply Planes},'' \emph{IEEE Transactions on
  Components, Packaging and Manufacturing Technology}, vol.~6, no.~8, pp.
  1268--1275, 2016.

\bibitem{FraunhoferTSV}
\BIBentryALTinterwordspacing
{Fraunhofer Institute for Reliability and Microintegration IZM}, ``{All Silicon
  System Integration Dresden ASSID - Fraunhofer IZM}.'' [Online]. Available:
  \url{https://www.izm.fraunhofer.de/en/abteilungen/\\
  /high{\_}density{\_}interconnectwaferlevelpackaging/ausstattung/\\
  /all{\_}silicon{\_}systemintegrationdresdenassid.html}
\BIBentrySTDinterwordspacing

\bibitem{Wu2017b}
J.~Wu, A.~Kodi, S.~Kaya, A.~Louri, and H.~Xin, ``{Monopoles Loaded with
  3-D-Printed Dielectrics for Future Wireless Intra-Chip Communications},''
  \emph{IEEE Transactions on Antennas and Propagation}, vol.~65, no.~12, pp.
  6838--6846, 2017.

\bibitem{CST}
\BIBentryALTinterwordspacing
``{CST Microwave Studio}.'' [Online]. Available: \url{http://www.cst.com}
\BIBentrySTDinterwordspacing

\bibitem{Lin2007}
J.~Lin, H.~Wu, Y.~Su, L.~Gao, A.~Sugavanam, and J.~Brewer, ``{Communication
  using antennas fabricated in silicon integrated circuits},'' \emph{IEEE
  Journal of Solid-State Circuits}, vol.~42, no.~8, pp. 1678--1687, 2007.

\bibitem{Timoneda2018a}
X.~Timoneda, S.~Abadal, A.~Cabellos-Aparicio, and E.~Alarc{\'{o}}n, ``{Modeling
  the EM Field Distribution within a Computer Chip Package},'' in
  \emph{Proceedings of the WCNC '18}, 2018.

\bibitem{Tasolamprou2018}
A.~C. Tasolamprou, M.~S. Mirmoosa, O.~Tsilipakos, A.~Pitilakis, F.~Liu,
  S.~Abadal, A.~Cabellos-Aparicio, E.~Alarc{\'{o}}n, C.~Liaskos, N.~V.
  Kantartzis, S.~Tretyakov, M.~Kafesaki, E.~N. Economou, and C.~M. Soukoulis,
  ``{Intercell wireless communication in software-defined metasurfaces},'' in
  \emph{Proceedings of the ISCAS '18}, 2018.

\bibitem{Ohira2011}
M.~Ohira, T.~Umaba, S.~Kitazawa, H.~Ban, and M.~Ueba, ``{Experimental
  characterization of microwave radio propagation in ICT equipment for wireless
  harness communications},'' \emph{IEEE Transactions on Antennas and
  Propagation}, vol.~59, no.~12, pp. 4757--4765, 2011.

\bibitem{Khademi2015}
S.~Khademi, S.~{Prabhakar Chepuri}, Z.~Irahhauten, G.~Janssen, and A.-J.
  van~der Veen, ``{Channel Measurements and Modeling for a 60 GHz Wireless Link
  Within a Metal Cabinet},'' \emph{IEEE Transactions on Wireless
  Communications}, vol.~14, no.~9, pp. 5098--5110, 2015.

\bibitem{Yan2009}
L.~Yan and G.~W. Hanson, ``{Wave propagation mechanisms for intra-chip
  communications},'' \emph{IEEE Transactions on Antennas and Propagation},
  vol.~57, no.~9, pp. 2715--2724, 2009.

\bibitem{Liu2015}
Y.~Liu, V.~Pano, D.~Patron, K.~Dandekar, and B.~Taskin, ``{Innovative
  propagation mechanism for inter-chip and intra-chip communication},'' in
  \emph{Proceedings of the WAMICON '15}, 2015.

\bibitem{Gade2017}
S.~H. Gade, S.~Garg, and S.~Deb, ``{OFDM Based High Data Rate, Fading Resilient
  Transceiver for Wireless Networks-on-Chip},'' in \emph{Proceedings of the
  ISVLSI '17}, 2017, pp. 483--488.

\bibitem{Yeh2013}
H.~H. Yeh, N.~Hiramatsu, and K.~L. Melde, ``{The design of broadband 60 GHz AMC
  antenna in multi-chip RF data transmission},'' \emph{IEEE Transactions on
  Antennas and Propagation}, vol.~61, no.~4, pp. 1623--1630, 2013.

\end{thebibliography}
\end{document}